\documentclass[conference]{IEEEtran}
\IEEEoverridecommandlockouts
\usepackage{cite}
\usepackage{amsmath,amssymb,amsfonts}
\usepackage{algorithmic}
\usepackage{graphicx}
\usepackage{textcomp}
\usepackage{xcolor}

\usepackage{enumitem}
\usepackage{booktabs} 
\usepackage{multirow}
\usepackage[table]{xcolor}

\def\BibTeX{{\rm B\kern-.05em{\sc i\kern-.025em b}\kern-.08em
    T\kern-.1667em\lower.7ex\hbox{E}\kern-.125emX}}
\begin{document}

\title{CATS: A Carbon-Aware Task Simulator for Reducing AI Data Center Emissions\\
}

\author{\IEEEauthorblockN{Dayuan Chen}
\IEEEauthorblockA{\textit{Computer Science Department} \\
\textit{Texas State University}\\
San Marcos, TX, USA \\
dayuan@txstate.edu}
\and
\IEEEauthorblockN{Ziliang Zong}
\IEEEauthorblockA{\textit{Computer Science Department} \\
\textit{Texas State University}\\
San Marcos, TX, USA \\
ziliang@txstate.edu}
}

\maketitle

\begin{abstract}
The rapid rise of generative AI is accelerating cloud data center expansion, with electricity demand projected to double by 2026. Because carbon-intensity varies by more than 5.5x across grids and times of day, where and when inference tasks execute significantly affects operational emissions. We address this issue with three aspects in this paper. First, we compile a global alignment dataset unifying 140 operational and planned cloud regions across 8 major providers with five-minute carbon-intensity traces for 145 grid regions from 2022 to 2024, revealing that 50\% of current sites lie in medium-to-high carbon-intensity grids, indicating a siting-carbon mismatch and unrealized carbon reduction potential. Second, we develop CATS (Carbon-Aware Task Simulator), a flexible trace-driven framework that profiles six AI inference tasks across multiple GPU types, synthesizes realistic diurnal curve, geographical and task mixes, and SLA constraints, and evaluates spatial and temporal schedulers against two baselines while reporting comprehensive metrics including carbon emissions, energy consumption, runtime, queue delay, and hardware utilization. Third, we quantify achievable $\text{CO}_2$ savings under realistic constraints: in a 24-hour trace with 600,000 tasks at fleet utilization of 0.37, spatial shifting reduces $\text{CO}_2$ by 38.4\% versus speed-first baseline, while temporal shifting yields 16\% savings with bounded SLA violations at 3.27\%. These results advocate locating future data centers in low carbon-intensity grids and demonstrate that carbon-aware scheduling on today's fleets can achieve substantial operational emissions reduction.
\end{abstract}


\section{Introduction}
Cloud computing has enabled AI applications to scale to trillion-query services within a few years, driving rapid adoption of GPU- and TPU-accelerated data centers (DCs).
In the United States alone, DC electricity use reached 176 TWh (4.4\% of national load) in 2023 and is expected to double or triple by 2028 as AI services multiply \cite{shehabi20242024, electricity2024analysis, ccam2025electricity}, with similar trajectories in Europe and Asia \cite{electricity2024analysis, ccam2025electricity, iea2025energyandai}. As GPUs and TPUs draw an order of magnitude more power than traditional CPU-powered servers, operational emissions from AI inference have become a pressing concern that cannot be ignored\cite{chen2025generative, tagarro2025forget, luccioni2024power}.

While a single AI inference requires far less computing and energy than model training, the volume of inference requests is massive. OpenAI revealed in early 2024 that about 100 billion words were generated every day. With an inference cost of \$30 per million tokens, their annual inference cost exceeds the estimated training cost of \$200 million\footnote{Inference cost for output only for gpt-4-0125-preview model, assuming 75 tokens per 100 words.}. This indicates that AI inference accounts for a major share of data center workload, energy use, and carbon emissions\cite{epochOptimallyAllocating}.

Operational emissions of an inference task are the product of its energy consumption and the carbon-intensity (CI) of the grid at the location and time where computation occurs. Two approaches exist to reduce these emissions: decreasing energy consumption or lowering carbon-intensity\cite{jegham2025hungry, patterson2021carbon}. Optimizing energy consumption includes improving the energy efficiency of all IT hardware and supporting infrastructure. Reducing CI can be achieved by strategically choosing locations in cleaner grid regions or through shifting tasks spatially and temporally to windows with lower grid emissions. This work focuses on CI reduction approaches for AI inference tasks.

While low-CI data center siting (i.e. placing DCs in regions with low average grid CI) has been studied in academic literature\cite{ayyildiz2025location}, the methodology remains largely theoretical. Industry DC providers often emphasize renewable energy procurement, but rarely disclose the actual CI of the grids supplying their DCs \cite{atmeta2024Sustainability, aboutamazon2024Amazon, apple2025envreport, microsoft2025Environmental, googlesustainability2025Environmental, schneider2024carbon}. Thus, a comprehensive, global-scale study on how existing data centers align with clean grids is essential but remains absent.

Workload shifting has also been presented, with several research proposing carbon-aware spatial and temporal scheduling across DCs\cite{radovanovic2022carbon, zhang2024carbon, souza2023casper, sukprasert2024limitations}. However, current approaches have several limitations. First, candidate data centers are often restricted to a single provider or limited to grids where CI data is available. Hardware capacity constraints are often overlooked, resulting in simplistic or absent modeling of queuing delay. Furthermore, many works do not consider realistic workload dynamics and arrival patterns. As a result, a comprehensive, flexible simulator for carbon-aware workload shifting that can model realistic hardware constraints, queuing behavior, and arrival patterns remains to be developed.

To address these gaps, we analyze the relationship between global grid carbon-intensity and existing as well as proposed data center locations, revealing how current deployments align with low-CI grids, and offering guidance for future carbon-aware site selection. In addition, we present CATS, a Carbon-Aware Task Simulator with plug-in shifting schedulers that supports any cloud providers, incorporates global grid carbon-intensity data, and enables flexible modeling of GPU types and hardware capacity constraints while respecting service level agreements (SLAs). CATS considers queuing delay, task mixes, and arrival patterns, and provides comprehensive reporting of key operational and sustainability metrics (e.g. carbon emissions, energy consumption, runtime, queuing delay, and GPU utilization) to enable detailed evaluation of carbon-aware task scheduling policies under realistic operational conditions.

Our paper makes three major contributions:
\begin{enumerate}
    \item We compile a multi-cloud siting-carbon alignment dataset by joining 140 cloud regions with five-minute marginal carbon-intensity traces for 145 grid regions from 2022 to 2024. We show that 50\% of existing sites reside in medium to high carbon-intensity grids, indicating a siting-carbon mismatch and highlighting opportunities for emission reduction.
    \item We develop the Carbon-Aware Task Simulator (CATS), which supports various AI inference tasks, multi-GPU profiling (runtime/energy), diurnal and geographically-skewed arrivals with task mix and SLAs. CATS is a discrete-event simulator that replays traces against two baselines (Speed-First and Carbon-First) and two carbon-aware policies (Spatial Shifting and Temporal Shifting).
    \item We demonstrate the effectiveness of CATS by quantifying the $\text{CO}_2$ savings and performance trade-offs under realistic constraints. In a 24-hour, 600,000-task trace at fleet utilization near 0.37, spatial shifting reduces $\text{CO}_2$ by 38.4\% relative to Speed-First, and temporal shifting reduces $\text{CO}_2$ by 16.1\% relative to Speed-First.
\end{enumerate}

The reset of the paper is organized as follows. Section II reviews related work on carbon-aware data center siting and design, carbon-intensity signals, sustainable AI, and task scheduling. Section III details our data sources and preprocessing pipeline and presents the siting-carbon alignment analysis across the U.S., EU (incl. UK), and Australia. Section IV details the CATS architecture and design. Section V provides experiments and evaluation on realistic traces across four schedulers. Section V concludes our study.

\section{Related Works}
\subsection{Carbon-Aware Data Center siting and Design}
Decisions of where to build data centers played a crucial role in determining their carbon footprints. A growing body of work studied where to build and how to design data centers to lower operational emissions. Ayyildiz proposed a multi-criteria decision framework that explicitly scored regions by renewable availability and grid factors to guide the siting of data centers \cite{ayyildiz2025location}; Similarly, Wang developed a comprehensive approach to optimize data center carbon emissions through siting and operational configuration \cite{wang2023carbon}; Al-Ayyoub modeled growth decisions under energy and cost constraints \cite{al2015optimizing}.
Other studies proposed integrated optimization of data center configuration and capacity shaping to minimize carbon output. Acun et al. integrated workload, grid signals and hardware choices to plan and operate data centers that maximized renewable energy usage and minimized carbon impact \cite{acun2023carbon}; Lin showed that dynamically adapting data center power usage and regional load shifting could raise renewable utilization \cite{lin2023adapting}; McMullen quantified and compared environmental improvements on site renewables for data centers \cite{mcmullen2024data}.

In industry, cloud providers committed to decarbonization \cite{atmeta2024Sustainability, aboutamazon2024Amazon, apple2025envreport, microsoft2025Environmental, googlesustainability2025Environmental}.
These pledges reinforced carbon-aware data center siting and design, but once a facility was built, it inherited the temporal variability of its grid and is tied to the local carbon profile, and these data center could not by themselves address the temporal variability of power generation or the needs of legacy data centers in carbon-intensive regions. These gaps motivated complementary operational solutions such as carbon-aware workload shifting to continuously optimize carbon efficiency after deployment, which is the focus of our study.

\subsection{Carbon-Intensity Data and Marginal Emissions Rates}
Operational decisions required accurate data about the electricity's carbon-intensity over time and location.
WattTime's ``Marginal Operating Emission Rate (MOER) Methodology'' \cite{watttimeMOER} formalized marginal signals for accurate real-time, location-aware accounting of IT workloads' carbon impact. MOER represented the incremental $\text{CO}_2$ emissions caused by an additional unit of power demand, contrasting with Average Operating Emission Rates (AOER), which simply divided total emissions by total generation across all resources, regardless of which plants are actually affected by a change in demand.
Unlike AOER, which served as a static attributional measure, MOER is a consequential indicator--it reflects the real-world causal impact of actions such as shifting demand or integrating renewables, because only the marginal plants adjust their output. MOER thus enabled accurate identification of optimal time periods when shifting workloads can directly lead to emission reductions, facilitating effective scheduling of flexible workloads to align consumption with periods of lower marginal emissions. Comparative studies showed that using AOER could mislead online scheduling, while MOER better captured real abatement potential \cite{sukprasert2024implications}. Empirical work demonstrated the preference of MOER for real-time control \cite{zhang2024carbon, jagannadharao2025timeshifting, chadha2023greencourier, dodge2022measuring, corradi2023marginal}. Our study adopts MOER for all scheduling and accounting.

\subsection{Sustainable Cloud Computing for AI}
Recent studies showed that cloud computing accounted for 2.5\% to 3.7\% of global CO2 emissions, which was greater than the global emissions created by commercial airline flights (roughly 2.4\%)\cite{overton2022issue, ferreboeuf2019lean}. The scale and growth of AI further pushed data center demand, making sustainability a real concern. Patterson et al. quantified training footprints and showed that careful choice of model, data center, and hardware had the potential to reduce carbon impact by 1000x. Analyses on real LLM deployments revealed that serving/inference could dominate energy consumption than model training, shifting the optimization from training alone to end-to-end or serving specific operations \cite{patterson2021carbon}. Everman showed that hardware and model selection and scheduling materially shifted serving-time emissions without impacting performance \cite{everman2023evaluating}.

\subsection{Carbon-Aware Workload Scheduling}
A mainstream research direction in sustainable computing was carbon-aware scheduling, which dynamically adjusted where and when workloads are executed to take advantage of cleaner energy.
Early pioneering studies demonstrated the potential of temporal shifting (delaying or advancing flexible task to greener times) and spatial shifting (routing tasks to greener geographic locations) approaches. For example, \cite{liu2011greening} showed that intelligently routing traffic across a set of distributed data centers could substantially increase the use of renewable energy. Qureshi et al. demonstrated the economic and energy upside of geographic shifting \cite{qureshi2009cutting}. Chien analyzed the carbon impact in AI inference, and showed the potential of geographical workload shifting \cite{chien2023reducing}.
On the temporal side, researchers developed schedulers that shifted deferrable batch tasks to off-peak or high-renewable periods. For example, Google's carbon-intelligent computing platform shifted daily non-urgent computing to times when each data center's local grid was cleaner \cite{radovanovic2022carbon}.\cite{lin2023adapting} explored adjusting data center capacity and load distribution in response to grid conditions, and found that regional load shifting combined with throttling could help integrate more renewable energy. Academic efforts generalized carbon-aware scheduling for various platforms: from cloud batch schedulers, web service, to inference serving \cite{asadov2025carbon, qi2024casa, chadha2023greencourier, kim2023greenscale}.

\subsection{Cloud Simulation Tool}
General purpose cloud simulation toolkits such as CloudSim \cite{calheiros2011cloudsim} provided a comprehensive framework to model and simulate cloud computing environments. While these simulators were flexible and widely used, they were not tailored to the carbon-aware scheduling or to the GPU-intensive characteristics of modern AI inference workloads.

In contrast to prior simulators, we introduce CATS, a trace-driven carbon-aware task simulator for cloud-scale AI inference. Our goal is to provide a reproducible framework for carbon-aware AI and data center operations that supports carbon-intensity signals, heterogeneous GPUs, multiple AI workloads, and pluggable scheduling policies. First, we quantify how today's multi-provider cloud footprint aligns with low-carbon grids by fusing public siting metadata with carbon-intensity traces and reporting siting distributions across carbon bands. Second, using CATS, we estimate the near-term decarbonization headroom without moving facilities by replaying realistic AI inference traces and comparing two baselines against carbon-aware spatial and temporal schedulers under capacity and delay constraints. Together, these results clarify where siting suffices, where it does not, and how much additional $\text{CO}_2$ reduction is achievable through operational workload shifting.

\section{Data Center Alignment Analysis}
In this section, we conduct a comprehensive global analysis using datasets including 140 operational and planned cloud regions across 8 major providers (e.g. Amazon AWS, Microsoft Azure, and Google Cloud etc.) with carbon-intensity dataset spanning 145 grid regions across the global from 2022 to 2024.

\subsection{Data Sources and Preprocessing}
\begin{figure}[htbp]
\centering
\includegraphics[width=1\linewidth]{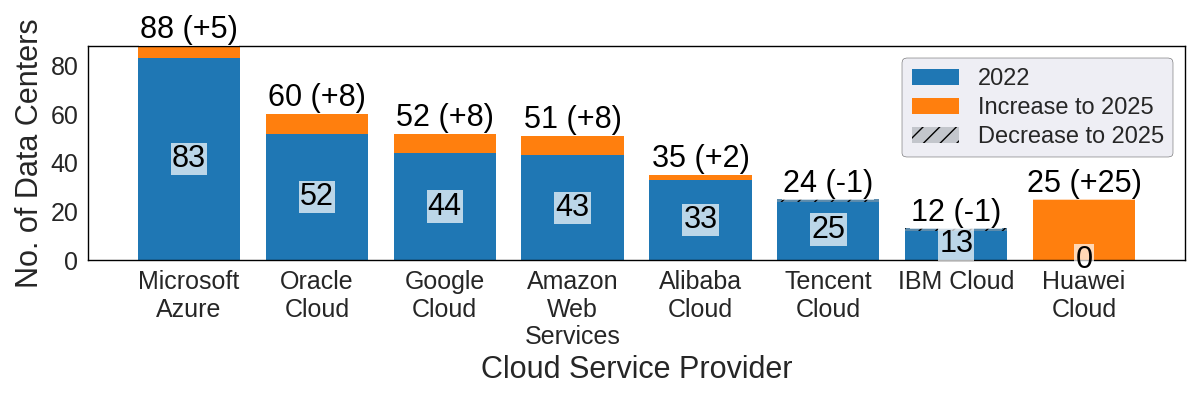}
\caption{\textbf{Cloud infrastructure by provider, 2022 vs 2025.} Bars show 2022 site counts (blue) and net change to 2025 (orange = increase; hatched = decrease). Labels give 2025 totals with the net change in parentheses. Based on TeleGeography’s Cloud Infrastructure Map, the combined number of DCs of the eight CSPs expanded from 293 in 2022 to 347 in 2025.}
\label{fig:cloud_infras_dist_inc}
\end{figure}

We collect cloud data center metadata from TeleGeography's Cloud Infrastructure Map \cite{CloudInfrastructureMap}, a widely used and publicly accessible resource that keeps tracks of global cloud infrastructure. For each site the dataset records the cloud service provider (CSP), the metro area (city, country), the number of availability zones (AZs), and an operational vs. planned flag. The dataset includes 347 data centers consist of 8 major CSPs representing 2025. For comparison, we also collected the 2022 snapshot from the same source, which contains 293 data centers from 7 CSPs. This three-year gap allows us to identify emerging data centers and analyze whether recent expansion aligns with low-carbon grids. Figure~\ref{fig:cloud_infras_dist_inc} summarizes provider counts and net changes. 

\begin{figure*}[htbp]
\centering
\includegraphics[width=\textwidth]{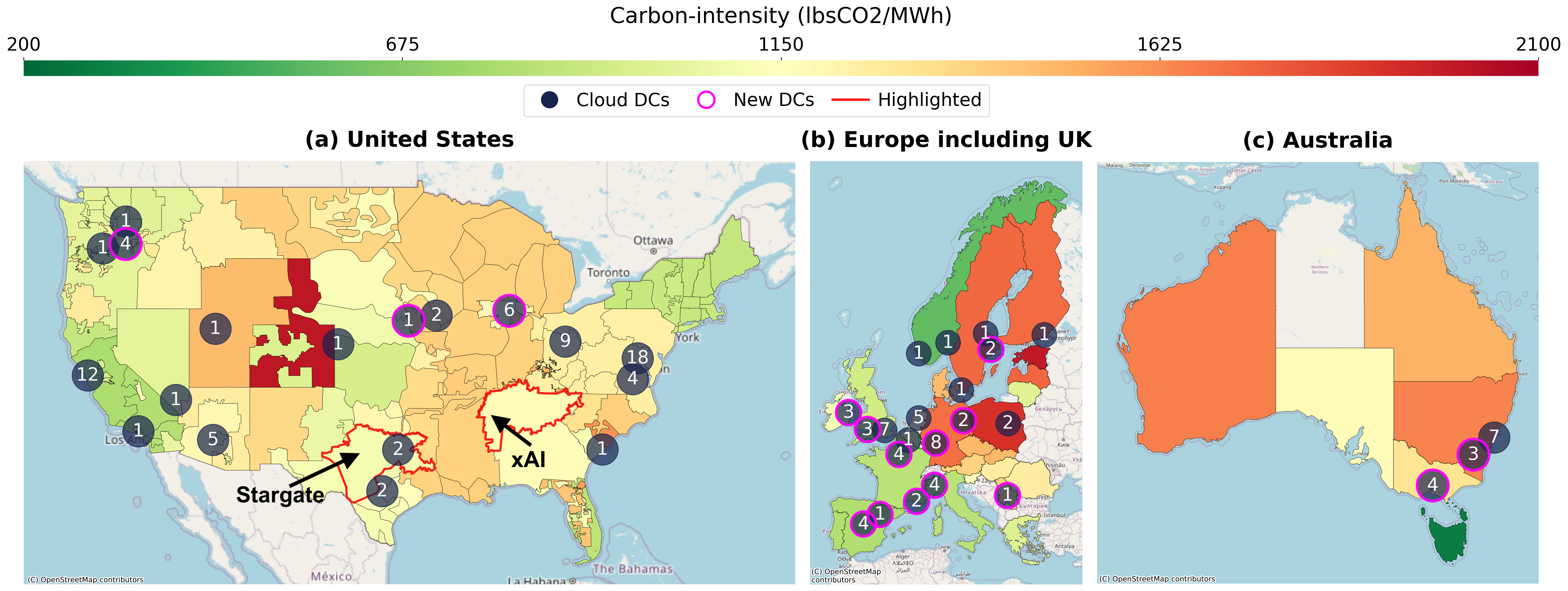}
\caption{Grids 2022-2024 average carbon-intensity with cloud data centers in three areas. The maps are divided into grid regions, the color in each region represents grid carbon-intensity, greener grids have lower carbon-intensity and hence less carbon footprint per energy consumption. The markers scattered across maps show the location and number of data centers, marker with colored circle indicate new data centers built since 2022. The two highlighted grids include the proposed data centers for Stargate Project (Abilene, TX) and xAI (Memphis, TN).}
\label{fig:moer_map_combined}
\end{figure*}

For carbon-intensity signal we use Marginal Operating Emission Rates (MOER) from WattTime \cite{watttimeMOER}, a recognized source employing empirical modeling approaches. MOER measures the extra carbon emissions generated caused by the additional electricity consumed in a grid (in $\text{lbsCO}_2\text{/MWh}$), which is the real-time emissions rate that changes every five minutes. The model comprehensively map grid load changes to various factor that directly affect local grid emissions, such as fuel mix, imports and exports, and renewable output, making it more representative for operational workload updates. This contrasts with the average operating emission rates, which simply relate total emissions with total grid generation and therefore unable to adapt scenarios that partially shifting the demand. Using MOER lets us better evaluate places and windows where executing flexible workloads directly reduces emissions.

We obtain five-minute MOER traces for 145 grid regions spanning the United States (114), Europe including the UK (25), and Australia (6), covering January 2022 to April 2025.  Throughout the paper, unless otherwise noted, ``carbon-intensity'' (CI) refers to MOER.

For global data center locations, we employ Geoapify Location Platform's geocoding API \cite{geoapify} to convert each data center metro area to latitude/longitude and map them to grid regions using WattTime's GeoJSON boundaries. For five-minute CI data, we convert the timestamp from grids' local time to UTC time for cross-region alignment. We address the only observed gap by forward-filling a missing window in WEM grid (Western Australia, from Nov. 7, 2024 22:00 to Nov. 8, 2024 00:30). The result dataset includes 122 cloud data centers in 2022 and 140 in 2025 with CI signals.

\begin{table*}[htbp]
\centering
\caption{Number of data centers in each CI range}
\begin{tabular}{r|rrrrrrrr|rrr|r}
\toprule
CI ($\text{lbsCO}_2\text{/MWh}$) & Azure & Oracle & Google & AWS & Alibaba & Tencent & IBM & Huawei & US & EU+UK & AUS & Total \\
\midrule
\textless{}900     &  8 &  7 &  4 &  5 & 2 & 2 & 1 & 0 & 12 & 17 &  0 & 29 \\
900-1200           & 13 & 13 &  9 &  2 & 1 & 0 & 2 & 1 & 22 & 19 &  0 & 41 \\
1200-1500          &  9 &  8 &  7 & 13 & 2 & 2 & 2 & 0 & 38 &  1 &  4 & 43 \\
\textgreater{}1500 &  8 &  5 &  5 &  4 & 2 & 1 & 2 & 0 &  0 & 17 & 10 & 27 \\
\midrule
Column Total              & 38 & 33 & 25 & 24 & 7 & 5 & 7 & 1 & 72 & 54 & 14 & 140 \\
\bottomrule
\end{tabular}
\label{tab:CI_CSP_count}
\end{table*}

\subsection{Alignment Analysis}
Figure~\ref{fig:moer_map_combined} visualizes available grids average carbon-intensity ($\text{lbsCO}_2\text{/MWh}$) from 2022 to 2024 and overlays the numbers of cloud data centers at each location in the United States, Europe, and Australia. Grids color from green to red shows their average CI. Greener grid regions indicate a cleaner grid and redder colored regions denote grids with more per unit carbon emissions. The markers scattered across maps show the number of data centers in this location, and for better visualization, data centers close to each other are combined and report in a single marker (distance threshold: 150 km for U.S. and Australia, 300 km for EU including UK). Markers circled by colored outline denote regions with at least one added data centers since 2022. Furthermore, we highlight two grids with red boundaries indicating two emerging data centers (the Abilene ``Stargate'' site in ERCOT\_NORTHCENTRAL in Texas and the Memphis xAI in TVA in Tennessee).

In all three maps, both existing data centers before 2022 and new builds since 2022 are not concentrated in low carbon regions.
In the United States (Fig.~\ref{fig:moer_map_combined}a showing 72 collected data centers), more data centers are on east and west coasts. There are 12 data centers located in the CAISO\_NORTH grid (approximately 748 $\text{lbsCO}_2\text{/MWh}$), representing about one-sixth of the total and situated in a relatively greener region with lower carbon intensity. In contrast, 22 are in PJM\_DC and 9 in PJM\_SOUTHWEST\_OH - both regions with much higher carbon intensity, around 1250 $\text{lbsCO}_2\text{/MWh}$. This means running the same workload in PJM grid would emits roughly 67\% more $\text{CO}_2$ than in greener west coast grids if identical energy are used. For data centers emerge since 2022, there are 4 in PACW (near Washington, CI: 1274), 2 in PJM\_CHICAGO (CI: 1190), and 1 in SPP\_SIOUX (in Iowa, CI: 1122). All of them have an average of over 1100 CI, which is not carbon efficient. The two newly built or under-construction large AI data centers - Stargate in Abilene, Texas (grid CI: 1,100) and another by xAI in Memphis, Tennessee (grid CI: 1,177) - are also not siting in low carbon intensity grids. 

In Europe including UK (Fig.~\ref{fig:moer_map_combined}b with 54 collected data centers), providers deploy across a wide CI spread rather than clustering in the cleanest grids. DE (Germany) and UK have most data centers (10 each, CI in DE: 1709, CI in UK: 929), followed by FR (France, 6, CI: 847). Of all data centers, there are 17 (31\%) located in regions with less than 900 CI, 18 (33\%) reside in over 1200 CI grids.
There are 17 new builds since 2022 across these area, 9 (53\%) below 900 CI, 1 between 900 and 1200 CI, and 7 located in regions with an average of more than 1200 CI (4 in ES, CI:812, 3 in FR:847, 2 in IT:850, 2 in UK:929, 1 in IE:1103, 1 in RS:1391, 2 in DE:1709, 1 in SE:1735, 1 in PL:1909). We can see in this area, providers tend to place new data centers in greener regions.

In Australia (Fig.~\ref{fig:moer_map_combined}c with 14 collected data centers), all facilities reside in two relatively carbon-intensive mainland regions (10 in NEM\_NSW with 1607 $\text{lbsCO}_2\text{/MWh}$, 4 in NEM\_VIC with 1236 $\text{lbsCO}_2\text{/MWh}$)). There are 2 new builds still in existing grids.

Overall, existing data centers and recent expansion patterns do not correct the carbon-siting mismatch. Table~\ref{tab:CI_CSP_count} classifies 140 facilities into four CI ranges for eight CSPs and three geographies (US, EU+UK, AUS). Using CI \textless{} 900 $\text{lbsCO}_2\text{/MWh}$ as ``low-carbon'', only 29 out of 140 (21\%) sites are in low-CI grids. 70 out of 140 (50\%) are in $\ge$ 1,200 side (medium-high to very high), and the remaining 41 out of 140 (29\%) fall in 900-1,200. By region, the U.S. tilts toward the 1,200-1,500 CI interval (39 sites) and no data center in very high CI grid; Europe places a substantial share in $\ge$ 1,500 (17 sites); Australia also skewed to \textgreater{} 1,500 (10 sites). By provider, no major CSP places a majority of its footprint in the \textless{} 900 range; each maintains sizable presence in $\ge$ 1,200 grids, with mix differences across providers but a shared pattern of weak low-CI concentration.

Two conclusions are derived from this global analysis. First, today's multi-provider footprint is misaligned with clean grids: half of sites in these three areas sit in $\ge$ 1,200 CI grids, while only around one-fifth are in $< 900$ CI grids. Second, recent growth has not improved this alignment, while Europe show an aware of siting over 50\% new data centers in low-CI grids. It is important for providers to take grid carbon-intensity into considerations when planning for future data centers. On the other hand, the results also motivate us to find solutions, under current data center siting, on carbon-aware scheduling across existing regions. In the following sections, we analyze the potentials of carbon reduction through workload shifting.

\section{CATS Design and Architecture} \label{sec:CATS}
In this section, we present the design and architecture of Carbon-Aware Task Simulator (CATS), an essential tool to support the evaluation of various carbon-aware scheduling algorithms under realistic operational conditions. CATS supports various AI inference task types, multi-GPU profiling, diurnal and geographically-skewed arrivals with task mix and per-task SLAs (delay limits).

CATS consists of five components: (1) a trace generator that synthesizes AI inference task traces; (2) a virtual cloud infrastructure that simulates data centers and GPU pools; (3) a discrete-event simulation engine that directs tasks to target resources; (4) a carbon accounting module that measures per-task carbon emissions at both scheduling time and execution time, and (5) a scheduling policy module that implements four scheduling policies. 

\subsection{Modeling the AI Inference Trace} \label{sec:trace_gen}

To model AI inference traces, CATS takes as input the benchmark runtime and energy data for each task–GPU pair. Users specify a few parameters: simulation start time (UTC), total duration (hours), number of tasks, task and region mix weights (each summing to 1), per-task delay limits (``HH:MM:SS''), and a 24-hour arrival curve (a vector of hourly weights). These settings can be defined in a single YAML file. Using this configuration and the benchmark table, CATS generates a reproducible, time-ordered inference trace. Each trace entry includes an arrival timestamp, task type, origin region, delay limit, and per-GPU mean runtime and energy values for scheduling.

Trace synthesis proceeds in four steps. First, the 24-hour diurnal vector is rotated to align with the simulation start hour, normalized over the simulation horizon, and used to allocate the total task count across hours. Second, within each hour the minute-level arrival counts are drawn from a Poisson distribution with mean equal to the hour's target divided by 60, and then adjusted to match the hourly total. Third, within each minute the tasks' arrival timestamps are sampled uniformly over the 60-second window, task types and origin regions are sampled independently from the normalized task mix and region mix lists, and each task is attached with its task-specific delay limit together with the per-GPU mean runtime and energy from the benchmark table. Fourth, all generated events are sorted by time, assign sequential event IDs, and written out as a CSV trace file.

To ensure reproducibility, we apply a deterministic seeding scheme: for each hour index $h$ the trace generator sets the random seed to the user-provided base seed plus $h$ before drawing Poisson minute counts and sampling task and region assignments. Given the same YAML configuration and base seed, the generator produces identical traces.

All timestamps in the trace are stored in UTC so the simulator can replay tasks in time order and align them with grid carbon-intensity signals. Because each trace records the per-GPU mean runtime and energy metrics, schedulers can later compare carbon impact and performance between devices without rerunning any profiling.

\subsection{Modeling the Cloud Infrastructure} \label{sec:cloud_model}
To model the cloud infrastructure, CATS requires the GPU capacity of each data center. This configuration is provided in the same YAML file used for trace generation configuration. CATS models the infrastructure with three levels: (i) multiple data centers; (ii) within each data center, one or more GPU pools (one pool per GPU type with a fixed number of identical devices); and (iii) single-GPU execution, where each task occupies exactly one device for its runtime.

Within each GPU pool, CATS maintains the total number of GPUs of that type, the number of devices currently running tasks, a normal queue to hold tasks when no free GPU available, a min-heap that tracks the predicted finish time of each device in the pool, a priority queue used by temporal scheduling policy to hold tasks that should run immediately when capacity becomes available, which can skip the normal queue, and a counter that tracks the total estimated runtime of tasks currently in the priority queue.

\begin{figure}[htbp]
    \centering
    \includegraphics[width=0.9\linewidth]{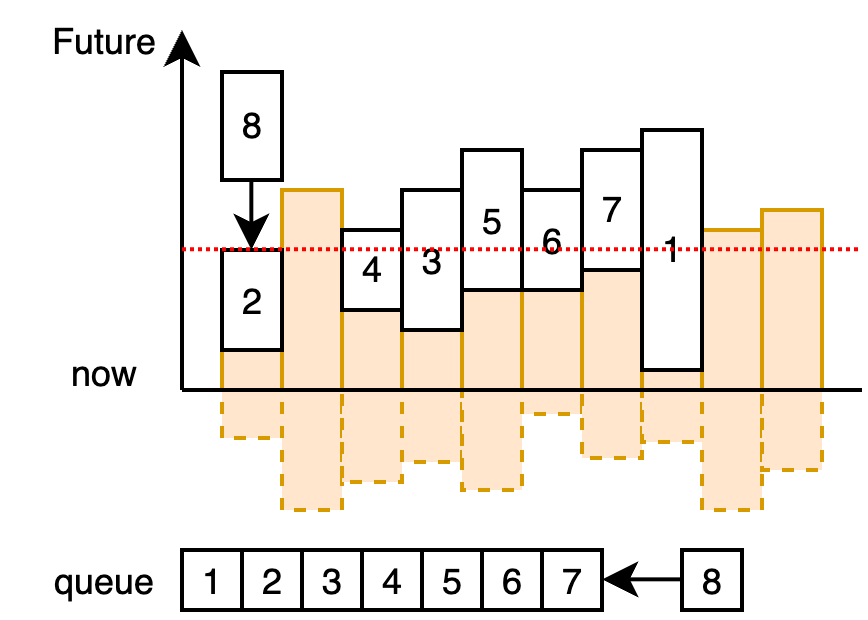}
    \caption{Task scheduling in a GPU pool. Bars cross horizontal axis indicate scheduled task on each GPU, the orange regions indicate current running tasks on each GPU. A min-heap is used to place incoming tasks to the earliest available GPU, as showed by numbered bars. The new task 8 is scheduled to the left most GPU, which will be executed after task 2 is completed.}
    \label{fig:scheduling_strategy}
\end{figure}

For fast predictions of start and finish times, every pool initializes a min-heap of ``next-free" times with one entry per device, set to the simulation start time. Assigning a task updates the entry who finishes earliest to the task's predicted finish time. Figure~\ref{fig:scheduling_strategy} illustrates this earliest-finish assignment within a pool. The orange bars cross horizontal axis indicate current running tasks on each GPU, and the numbered white bars show the already scheduled tasks on each GPU, the numbers denote the order of tasks were scheduled when the scheduler prioritizes task latency. When a new task (task 8) arrives at this GPU pool, it will be placed on the device whose finish time is the smallest in the min-heap, so that it starts immediately after task 2 completes.

\subsection{Scheduling Engine and Workflow}
We implement a customized, trace-driven discrete event simulator (DES) as the scheduling engine to orchestrate between the arrival trace, the cloud infrastructure model, the scheduling policy, and the carbon accounting module. The DES consumes the synthetic trace in time order, maintains a global event heap, and repeatedly pops the next event until no events remain. For each event, it queries the scheduler for placement decisions, updates pool capacities and queues while enforcing constraints, logs per-task carbon and performance metrics, and outputs a trace where each task has its actual start time, end time, and assigned GPU pool. Given a fixed trace, configuration, and scheduling policy, the engine is fully deterministic.

\begin{figure}[htbp]
    \centering
    \includegraphics[width=0.9\linewidth]{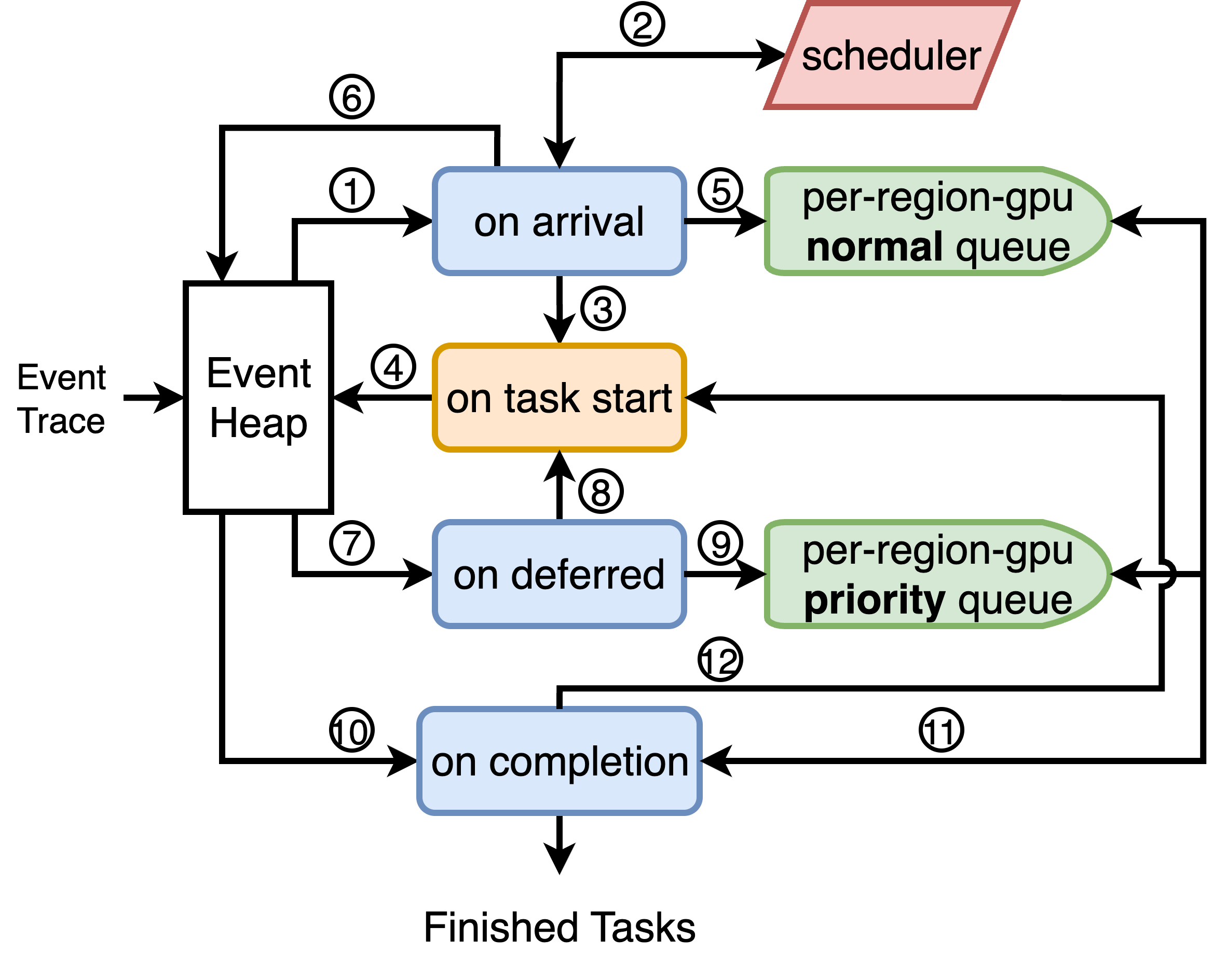}
    \caption{Carbon-Aware Task Simulator (CATS) Scheduling Engine workflow.}
    \label{fig:des}
\end{figure}

The synthetic trace is kept in a min-heap over time and each is labeled with an event type, initialized to ARRIVAL event. The event calendar contains three event types: ARRIVAL, DEFERRED, and COMPLETION, as illustrated in Figure~\ref{fig:des}. On arrival {\large \textcircled{\small 1}}, the simulator calls the scheduler with the task, a snapshot of all data-center states, and current time. The scheduler selects a target region and GPU pool and returns either an immediate placement or a deferred placement within the task's delay limit {\large \textcircled{\small 2}}. For an immediate placement, the DES refreshes the target pool's min-heap of ``next-free'' times and, if a device is available it starts the task immediately {\large \textcircled{\small 3}}, updates task predicted finish time, and pushes a completion event at the at the predicted finish time {\large \textcircled{\small 4}}; otherwise, the task joins the pool's normal queue {\large \textcircled{\small 5}}. For a deferred decision, the simulator inserts a DEFERRED event for this task at the requested release time {\large \textcircled{\small 6}}. 

When a DEFERRED events fires {\large \textcircled{\small 7}}, the task should run as soon as possible to keep the carbon benefits. The DES again checks the target pool: if a GPU is idle it starts the task immediately {\large \textcircled{\small 8}}; otherwise, the task is placed in the pool's priority queue {\large \textcircled{\small 9}}. Both normal and priority queues are first-in-first-out (FIFO), but when capacity frees, the simulator looks at the head of each queue and applies an earliest-deadline-first rule so that deferred tasks with earlier deadlines can cut ahead of non-deferred tasks while never preempting a running task.

On completion {\large \textcircled{\small 10}}, the simulator frees one GPU in the corresponding pool, triggers the carbon accounting module to compute the task's average carbon-intensity and emissions over its execution window, and updates performance statistics. It then checks the pool's queues: if there are waiting tasks, it selects the next one using the same earliest-deadline rule {\large \textcircled{\small 11}}, and starts this task immediately {\large \textcircled{\small 12}}, then scheduling a new COMPLETION event. Throughout the run, the DES logs per-task events and scheduler decision, per-pool state changes for audits.

\subsection{Scheduling Policies} \label{sec:scheduling_policies}
\begin{figure}[htbp]
    \centering
    \includegraphics[width=0.9\linewidth]{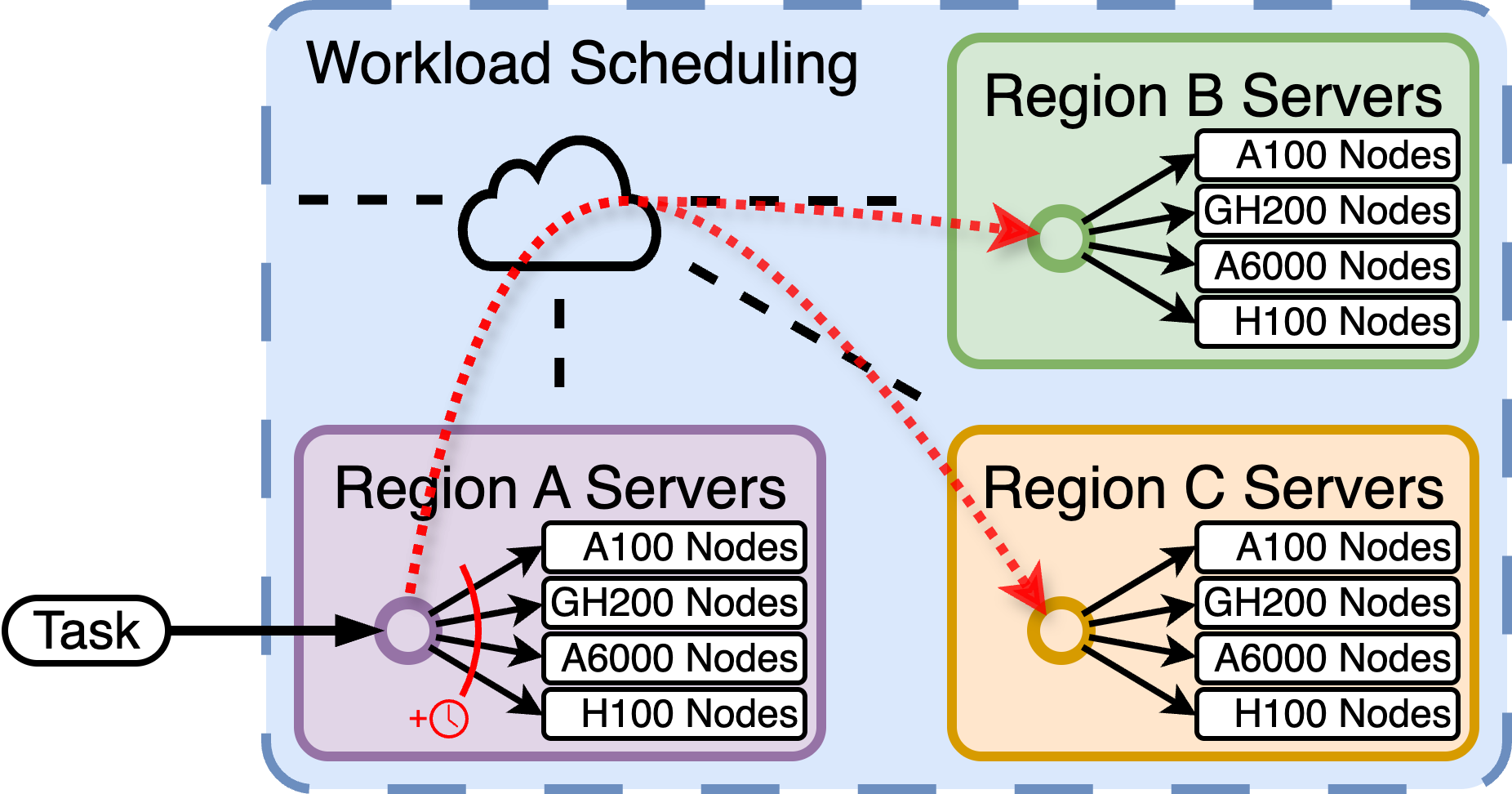}
    \caption{Workload shifting illustration. A task is initially sent to region A servers, spatial shifting redirects the task to another region while temporal shifting defers the task for execution.}
    \label{fig:ls_scope}
\end{figure}

Our goal is to minimize the operational $\text{CO}_2$ by making scheduling decisions that exploit variation in carbon-intensity across space and time. Within CATS, scheduling policies are implemented as pluggable modules that share the same interface: on each task arrival the scheduler observes the current data center state, chooses where to run (spatial shifting across data centers) or when to start (temporal shifting at local data center), and GPU type selection when multiple node types are available, while respecting capacity and SLA constraints, and return either a immediate placement decision or a deferral placement decision. Figure~\ref{fig:ls_scope} illustrates the general view of this workload scheduling.
We implement Speed-First and Carbon-First as two baseline scheduling algorithms, and Spatial Shifting and Temporal Shifting as two carbon-aware workload shifting algorithms.

\textbf{Speed-First baseline policy.}
In the speed-first baseline ($baseline_{speed}$), tasks always stay at their origin data center. At each task arrival, the scheduler ranks GPU types by predicted finish time, computed as each pool’s next available time plus the task’s mean runtime on that GPU. It selects the pool with the earliest predicted finish, breaking ties by shorter runtime. If the chosen pool has an idle GPU, the task starts immediately; otherwise, it is queued, and the pool’s min-heap is updated accordingly.

\textbf{Carbon-First baseline policy.}
The carbon-oriented baseline ($baseline_{carbon}$) also keeps tasks in their origin region. For each GPU type there, it predicts the task’s start and finish window based on the pool’s queue, then estimates emissions by multiplying the region’s carbon intensity over that window by the task’s mean energy. It selects the option with the lowest predicted emissions, breaking ties by earlier finish time. If no option meets the delay limit, it falls back to the lowest-emission choice even if this causes an SLA violation. As before, tasks either start immediately on an available GPU or join the normal queue.

\begin{figure}[htbp]
    \centering
    \includegraphics[width=0.99\linewidth]{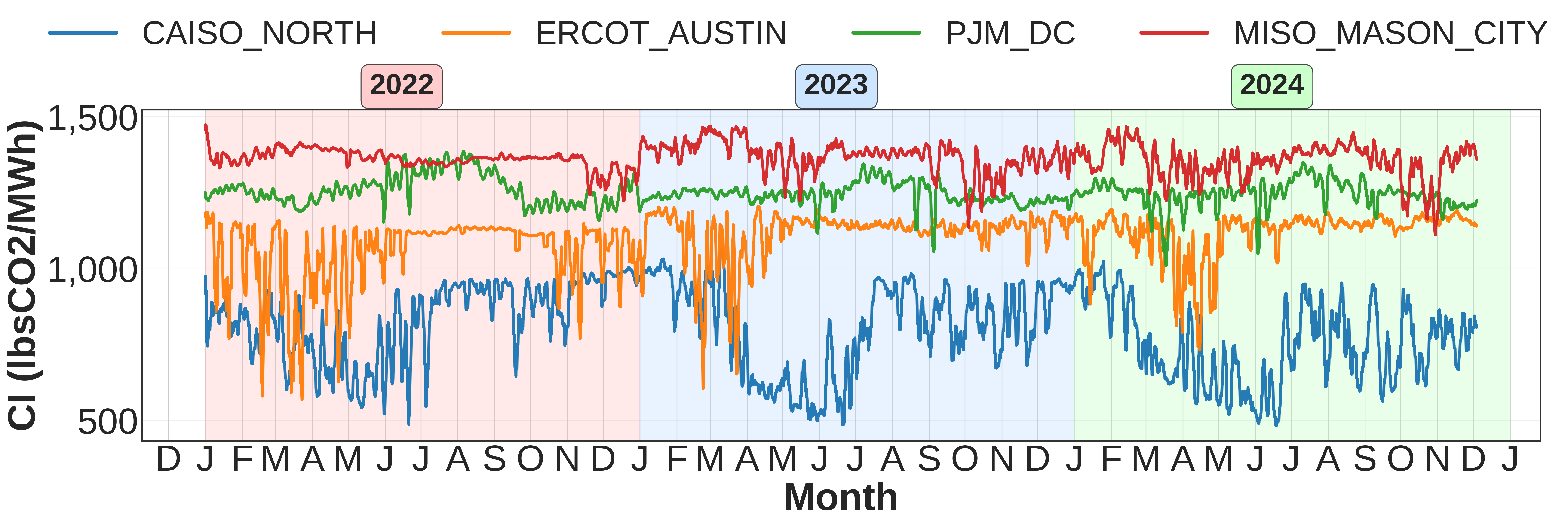}
    \caption{Carbon-intensity (CI) Trends Across Selected Regions. Three-year carbon-intensity (3-day moving average) for four U.S. regions. Persistent inter-regional gaps indicate substantial headroom for carbon-aware spatial shifting.}
    \label{fig:spatial_moer}
\end{figure}

\textbf{Spatial shifting policy.}
The spatial shifting policy extends the Carbon-First baseline by allowing routing across data centers. For each region-GPU candidate, it predicts a start/finish window using that pool's queue, computes the associated emissions, and selects the lowest-emission candidate that respects the task's delay limit. To avoid shifting for negligible gains, it applies minimum relative and absolute saving gates: by default it only shifts away from the origin if the best remote option reduces predicted emissions by at least $\epsilon_{s,rel}=0.05$ (5\%) and by more than a small absolute threshold. If the best origin option has zero predicted emissions (e.g. during a zero-carbon window) or no SLA-safe remote candidate exists, the task stays local; if neighter local nor remote can meet the delay limit, the scheduler return the best origin option with SLA violation. Figure~\ref{fig:spatial_moer} shows persistent inter-regional carbon-intensity gaps that create opportunities for spatial shifting.

\begin{figure}[htbp]
    \centering
    \includegraphics[width=0.99\linewidth]{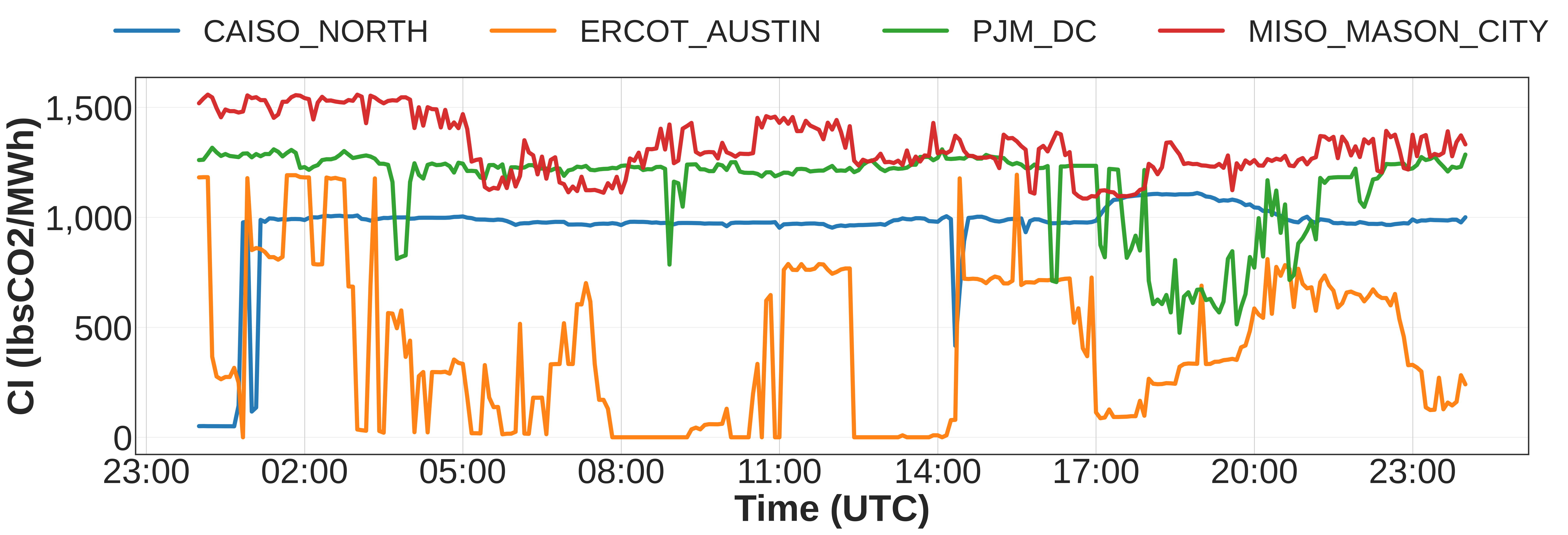}
    \caption{Intraday carbon-intensity over a 24-hour period for four regions. Pronounced within-day swings enable carbon savings via bounded temporal shifting at the original site.}
    \label{fig:daily_moer}
\end{figure}

\textbf{Temporal shifting policy.} The temporal shifting policy keeps tasks in their origin region but searches within each task’s delay limit for a future release time that minimizes predicted emissions. For each local GPU type, it predicts earliest queue-aware start time, scans candidate start times up to the latest SLA-safe release, and evaluates emissions for each start/finish window. A candidate is admissible only if it (i) passes a rate cap $\alpha$ limiting how many GPU-seconds of work can be newly deferred out of the current five-minute window of the pool $(d,g)$, (ii) satisfies a capacity ledger $\rho$ that caps the deferred GPU-seconds landing in each future per-minute bucket over a 24-hour horizon, and (iii) respects a backlog cap $\beta$ bounding the total deferred GPU-seconds across that horizon. These three guards are maintained independently for every region-GPU pair. The policy also enforces minimum relative and absolute saving gates (e.g., $\epsilon_{t,rel}=0.05$ and $\epsilon_{t,abs}=0.5\ (\text{gCO}_2)$) to avoid deferring for trivial savings. If a candidate passes the savings gates and all three guards, the scheduler returns a ``defer-until" decision. The DES will then places the task at the chosen release time, using the priority queue with earliest-deadline-first between queues to protect SLAs. If no admissible deferral is found, the policy falls back to the previous Carbon-First placement at arrival. Figure~\ref{fig:daily_moer} illustrates the intraday carbon-intensity swings that enable temporal shifting.

\subsection{Carbon Emission Accounting}
CATS computes operational emissions by aligning each task's execution window with the carbon-intensity series of its execution region and multiplying by the task's measured energy. Let $\mathcal{D}$ be the set of data centers. Each data center $d \in \mathcal{D}$ maps to a grid region with carbon-intensity $\text{CI}_d(t)$ ($\text{lbsCO}_2/\text{MWh}$) sampled every five minutes. Data center $d$ supports one or more GPU pools, each pool contains a fixed number of identical devices of a given GPU types.

Tasks arrive as a fixed trace $\mathcal{J}$. A task $j \in \mathcal{J}$ has arrival time $a_j$, origin data center $o_j$, a task type, and per-GPU benchmarks: mean runtime $r_{j,g}$ (seconds) and mean energy $E_{j,g}$ (joules) for each GPU type $g$. Each task has a delay limit $\Delta_j$ and is non-preemptive, occupying exactly one GPU for its entire runtime.

The system parameters used in this subsection are summarized in Table~\ref{tab:notation}.
\begin{table}[htbp]
\caption{Notation and Parameters}
\label{tab:notation}
\begin{center}
\begin{tabular}{ll}
\toprule
\textbf{Notation} & \textbf{Description} \\
\midrule
\multicolumn{2}{l}{\textit{System Configuration}} \\
$\mathcal{D}$ & Set of data centers \\
$\text{CI}_d(t)$ & Carbon-intensity signal in data center $d$ at time $t$ \\
$N_{d,g}$ & Number of GPUs of type $g$ at data center $d$ \\
\midrule
\multicolumn{2}{l}{\textit{Workload Characteristics}} \\
$a_j$ & Arrival time of task $j$ \\
$\Delta_j$ & Delay limit of task $j$ \\
$s_j$ & Predicted start time of task $j$ \\
$d_j$ & Selected data center for task $j$ \\
$g_j$ & Selected GPU pool for task $j$ \\
$r_{j,g}$ & Mean service time (seconds) for task $j$ on GPU type $g$ \\
$E_{j,g}$ & Mean energy consumption for task $j$ on GPU type $g$ \\
\bottomrule
\end{tabular}
\end{center}
\end{table}

On arrival, the scheduler finds the optimal choice and returns the selected data center $d_j$, GPU pool $g_j$, a flag indicating whether the task can start immediately or should enqueue, and, for the temporal policy, whether the task should be deferred.
The choice is determined by evaluating the task's predicted start time $s_j$ at each candidate $(d,g)$ ($s_j$ is candidate-specific, i.e., $s_{j,(d,g)}$), accounting for queuing delay (for non-temporal policies), or by searching $(s_j, g)$ pairs, the potential release times at each local GPU pool (temporal policy), and computing the predicted carbon emissions, the product of its energy on that GPU pool (converted to MWh) and the time-weighted average carbon-intensity in its execution region over task's runtime. For a candidate $(d,g)$ with predicted window $[s_j, s_j + r_{j,g}]$, we estimate:
\begin{equation}
    \text{CO}^{pred}_2(j;d,g) = (\frac{E_{j,g}}{3.6\times 10^9}) 
    \frac{1}{r_{j,g}}
    \int_{s_j}^{s_j+r_{j,g}}\text{CI}_d(t) dt
\end{equation}
where $\text{CI}_d$ is in $\text{lbsCO}_2/\text{MWh}$
and $E_{j,g}$ is in joules. The factor $3.6 \times 10^{9}$ converts joules to MWh so the results is in pounds of $\text{CO}_2$. In practice, the integral is evaluated by summing over the five-minute carbon-intensity signals that overlap the task's runtime, this proportionally accounts for partial windows at both ends.

When determining the start time $s_j$, the scheduler tries to satisfy:
\begin{equation}
    s_j = \text{max}\left\{ a_j, \text{next\_free}(d, g) \right\}
\end{equation}
\begin{equation}
    s_j + r_{j,g} \le a_j + \Delta_j
\end{equation}
If no candidate satisfies the delay limit boundary, the scheduler falls back to the best available option and records an SLA violation. For temporal shifting policy, at placement time, the scheduler estimates the start time using the earliest free GPU from the min-heap and, if priority queue is not empty, it adds an aggregate correction $s_j \mathrel{+}= prio\_q\_runtime\ /\ N_{d,g}$. This adjustment averages the priority work across the pool and therefore is an approximation of how priority and normal queues will actually interleave. And this indeterminacy on which task start first may cross its deadline even though both looked safe at placement, which means the realized start time can exceed $s_j$, we record this as prediction error.

Our objective is to minimize total operational emissions:
\begin{equation}
    min\sum_{j \in \mathcal{J}} \text{CO}_2(j)
\end{equation}
Decision-time comparisons use $\text{CO}_2^{pred}$ as above, while actual accounting at completion replaces $[s_j, s_j+r_{j,g}]$ with the realized start and finish time.

This optimization is subject to three hard constraints enforced by the schedulers: (i) Capacity: at any time, the number of concurrent active tasks in pool $(d,g)$ never exceeds its pool capacity; (ii) SLA/delay limit: for all tasks, $s_j + r_{j,g} \le a_j + \Delta_j$. (except when a policy explicitly allows SLA violation as a fallback); and (iii) Admissible regions: each task may be restricted to a subset $A_j \subseteq \mathcal{D}$. In our experiments, $A_j = {o_j}$ for both baselines and the temporal scheduler, and $A_j = \mathcal{D}$ for the spatial scheduler.

\section{Experiments and Evaluation}
In this section, we use CATS to evaluate four scheduling policies on realistic AI inference workloads. We first profile six types of AI inference tasks across four different GPUs, then synthesize a 24-hour trace with configurable task mix, region mix, and diurnal pattern, and finally replay the trace through the discrete-event simulator described in Section~\ref{sec:CATS} using two baselines and two carbon-aware shifting policies.

\subsection{AI Inference Task Profiling}
We profile six widely used AI inference tasks types: text generation, text-to-speech, text-to-image, image captioning (image-to-text), image-to-image, and text-to-video. For each task family, we select multiple open-source models hosted on Hugging Face to span different compute footprints (e.g., three text-generation models from 2.7B to 13B parameters, lightweight and XL variants for diffusion models, and two text-to-video models).
Table~\ref{tab:task_model} lists the models and their aliases.

\begin{table}[htbp]
\caption{AI Inference Tasks and Models used}
\label{tab:task_model}
\begin{center}\resizebox{1.0\columnwidth}{!}{
\begin{tabular}{lll}
\toprule
Task type & Model & Alias \\
\midrule
\multirow{3}{*}{Text Generation} & meta-llama/Llama-2-13b-chat-hf & TG--Llama \\
                                 & mistralai/Mistral-7B-Instruct-v0.3 & TG--Mistral \\
                                 & microsoft/phi-2 & TG--phi2 \\
\midrule
\multirow{3}{*}{Text to Speech}  & suno/bark & TTS--bark \\
                                 & speechbrain/tts-tacotron2-ljspeech \& & \multirow{2}{*}{TTS--tacotron2} \\
                                 & \, speechbrain/tts-hifigan-ljspeech \\
\midrule
\multirow{2}{*}{Text to Image}   & stabilityai/stable-diffusion-xl-base-1.0 & T2I--sdxl \\
                                 & OFA-Sys/small-stable-diffusion-v0 & T2I--small \\
\midrule
\multirow{2}{*}{Image to Text}   & Salesforce/blip2-flan-t5-xl & I2T--blip2 \\
                                 & nlpconnect/vit-gpt2-image-captioning & I2T--vigpt2 \\
\midrule
\multirow{2}{*}{Image to Image}  & stabilityai/stable-diffusion-xl-base-1.0 & I2I--sdxl \\
                                 & OFA-Sys/small-stable-diffusion-v0 & I2I--small \\
\midrule
\multirow{2}{*}{Text to Video}   & damo-vilab/text-to-video-ms-1.7b & T2V--ms \\
                                 & cerspense/zeroscope\_v2\_576w & T2V--zeroscope \\
\bottomrule
\end{tabular}
}\end{center}
\end{table}

All models are deployed on four types of GPUs in the cloud: A100, A6000, H100, and GH200. Detailed host and device specifications are shown in Table~\ref{tab:HW_spec}. Each task is executed on a single GPU.

\begin{table}[htbp]
\caption{Hardware Specification for Benchmarks}
\label{tab:HW_spec}\begin{center}\resizebox{1.0\columnwidth}{!}{%
\begin{tabular}{ccc}
\toprule
Node & Host  Specification & GPU Information \\ \midrule
NVIDIA  A100 &
\begin{tabular}[c]{@{}c@{}}AMD EPYC 7J13\\ 30 vCPU \& 220 GB RAM\\ 512 GB SSD\end{tabular} &
\begin{tabular}[c]{@{}c@{}}1 x A100\\ 40GB SXM4\\ Ampere\end{tabular} \\
\midrule
NVIDIA  GH200 &
\begin{tabular}[c]{@{}c@{}}Neoverse-V\\ 64 vCPU \& 432 GB RAM\\ 4 TB SSD\end{tabular} &
\begin{tabular}[c]{@{}c@{}}1 x GH200\\ 96GB/480GB\\ Hopper\end{tabular} \\
\midrule
NVIDIA  A6000 &
\begin{tabular}[c]{@{}c@{}}AMD EPYC-Rome\\ 14 vCPUs \& 100 GB RAM\\ 512 GB SSD\end{tabular} &
\begin{tabular}[c]{@{}c@{}}1 x A6000\\ 48GB\\ Ampere\end{tabular} \\
\midrule
NVIDIA  H100 &
\begin{tabular}[c]{@{}c@{}} Intel(R) Xeon(R) Platinum 8480+\\ 26 vCPU \& 225 GB RAM\\ 1 TB SSD\end{tabular} &
\begin{tabular}[c]{@{}c@{}}1 x H100\\ 80GB PCIe\\ Hopper\end{tabular} \\
\bottomrule
\end{tabular}}\end{center}
\end{table}

\renewcommand{\arraystretch}{1.1}
\begin{table*}[htbp]
\caption{AI Inference Task Benchmark}
\label{tab:task_profiles}
\begin{center}
\rowcolors{2}{gray!20}{white}
\begin{tabular}{l|rrrr|rrrr|rrrr}
\toprule
\rowcolor{white}
\multirow{2}{*}{Task Type} 
& \multicolumn{4}{c}{Runtime (s)} 
& \multicolumn{4}{|c}{Energy (J)} 
& \multicolumn{4}{|c}{Power (W)} \\
\cmidrule(lr){2-5} \cmidrule(lr){6-9} \cmidrule(lr){10-13}
\rowcolor{white}
& \multicolumn{1}{c}{A100} 
& \multicolumn{1}{c}{A6000} 
& \multicolumn{1}{c}{GH200} 
& \multicolumn{1}{c|}{H100}
& \multicolumn{1}{c}{A100} 
& \multicolumn{1}{c}{A6000} 
& \multicolumn{1}{c}{GH200} 
& \multicolumn{1}{c|}{H100}
& \multicolumn{1}{c}{A100} 
& \multicolumn{1}{c}{A6000} 
& \multicolumn{1}{c}{GH200} 
& \multicolumn{1}{c}{H100}  \\
\midrule
TG--Llama       & 8.15 & 10.18 & 7.50 & \textbf{5.09} 
                & 1,549.50 & 2,915.87 & 2,518.05 & \textbf{1,133.20} 
                & \textbf{190.12} & 286.56 & 335.63 & 222.69 \\
TG--Mistral     & 6.23 & 7.53 & 5.82 & \textbf{3.39} 
                & 903.01 & 1,748.79 & 1,728.90 & \textbf{636.61} 
                & \textbf{144.96} & 232.12 & 296.81 & 187.48 \\
TG--phi2        & 4.98 & 5.64 & 4.35 & \textbf{2.85} 
                & 514.04 & 917.78 & 1,139.53 & \textbf{370.43} 
                & \textbf{103.23} & 162.82 & 261.68 & 130.15 \\
TTS--bark       & 39.90 & 42.23 & 34.69 & \textbf{17.52} 
                & 2,747.55 & 4,913.62 & 8,802.31 & \textbf{2,243.61} 
                & \textbf{68.85} & 116.35 & 253.74 & 128.07 \\
TTS--tacotron2  & 0.98 & 1.10 & \textbf{0.89} & 0.93 
                & \textbf{63.18} & 113.32 & 221.42 & 91.20 
                & \textbf{64.30} & 103.43 & 249.22 & 97.94 \\
T2I--sdxl       & 8.94 & 14.67 & 7.10 & \textbf{6.35} 
                & 3,154.43 & 4,276.16 & 3,506.97 & \textbf{2,022.75} 
                & 352.93 & \textbf{291.45} & 493.91 & 318.53 \\
T2I--small      & 1.72 & 1.99 & 1.47 & \textbf{1.05} 
                & 294.25 & 556.71 & 572.77 & \textbf{285.89} 
                & \textbf{228.78} & 279.82 & 390.78 & 272.45 \\
I2T--blip2      & 0.32 & 0.41 & 0.34 & \textbf{0.16} 
                & 30.68 & 60.78 & 95.56 & \textbf{23.27} 
                & \textbf{96.01} & 149.79 & 279.66 & 147.54 \\
I2T--vigpt2     & 0.15 & 0.43 & \textbf{0.11} & 0.12 
                & \textbf{9.90} & 42.72 & 30.28 & 13.58 
                & \textbf{66.77} & 100.22 & 263.11 & 109.47 \\
I2I--sdxl       & 5.52 & 5.80 & 4.58 & \textbf{2.63} 
                & 370.58 & 622.81 & 1,165.91 & \textbf{301.94} 
                & \textbf{67.17} & 107.47 & 254.31 & 114.82 \\
I2I--small      & 1.21 & 1.28 & 1.05 & \textbf{0.66} 
                & 121.15 & 177.55 & 293.52 & \textbf{97.87} 
                & \textbf{99.86} & 138.52 & 280.36 & 148.09 \\
T2V--ms         & 442.86 & 484.99 & 352.22 & \textbf{197.14} 
                & 26,173.68 & 43,160.25 & 83,512.17 & \textbf{19,078.01} 
                & \textbf{59.10} & 88.99 & 237.10 & 96.77 \\
T2V--zeroscope  & 799.59 & 882.83 & 693.04 & \textbf{324.04} 
                & 57,316.20 & 86,561.45 & 166,077.96 & \textbf{34,020.76} 
                & \textbf{71.68} & 98.05 & 239.64 & 104.99 \\
\bottomrule
\end{tabular}
\end{center}
{\raggedright Note: All results in this table are benchmarked on actual GPUs and models. \par}
\end{table*}

Within each task family, we fix the inputs across models to make comparisons meaningful.
For tasks requiring text input (text generation, text-to-speech, text-to-image, and text-to-video), the input prompts average 29 tokens\footnote{Token counts are collected using the OpenAI tokenizer for GPT-4o and GPT-4o mini at https://platform.openai.com/tokenizer.}.
For image-based tasks (image-to-text and image-to-image), we use a single 224$\times$224 color JPEG image (25.7 KB). The image-to-image task additionally includes a guiding text prompt.
Outputs are stored in their native formats: text generation and image captioning results are logged as text in JSON; text-to-speech outputs are saved as WAV files (22.05 kHz for Tacotron2, 24 kHz for Bark); text-to-image and image-to-image outputs are saved as PNGs; and text-to-video results are exported as MP4 files at 16 fps. CATS records only runtime and device energy for analysis.

For each model–GPU pair, we perform two warm-up iterations followed by ten measured runs, recording end-to-end runtime and on-device energy via NVML. We average the ten runs and compute mean power as energy divided by runtime. Results in Table~\ref{tab:task_profiles} indicate that text-to-video workloads are the most compute- and energy-intensive, while image captioning, image-to-image, smaller text-generation models, and Tacotron2 are the fastest and least energy-demanding. Hopper devices (H100/GH200) generally outperform Ampere devices (A100/A6000) in speed. H100 consumes the least energy for most tasks, whereas A100 often achieves the highest power efficiency. In subsequent experiments, the per-pair mean runtime is used as the job “size” signal, and mean energy is used for emission accounting in combination with carbon-intensity time series.

\subsection{Workload and Trace Configuration}
We generate a 24-hour trace with $N=600{,}000$ tasks using the trace generator described in Section~\ref{sec:trace_gen}. Since detailed production mixes are rarely public, we adopt a plausible composition dominated by text generation: over 60\% of tasks are text-generation, 15\% are text-to-speech, and the remainder are distributed across image and video tasks, with larger models favored within each category. Table~\ref{tab:task_mix} summarizes the per-model shares and per-task delay limits. Small tasks have tight deadlines, while larger tasks, such as text-to-video, are allowed to be deferred for up to three hours.

\begin{table}[htbp]
\caption{Task Mix and Defer Limit}
\label{tab:task_mix}
\begin{center}
\begin{tabular}{lrr}
\toprule
Task-Model Alias & Share ($\sum=1.0$) & Delay limit (HH:MM:SS) \\
\midrule
TG--Llama        & 0.28 & 00:03:30 \\
TG--Mistral      & 0.28 & 00:02:30\\
TG--phi2         & 0.06 & 00:02:00 \\
\midrule
TTS--bark        & 0.10 & 00:15:00 \\
TTS--tacotron2   & 0.05 & 00:00:30 \\
\midrule
T2I--sdxl        & 0.08 & 00:05:00 \\
T2I--small       & 0.03 & 00:01:00 \\
\midrule
I2T--blip2       & 0.04 & 00:00:30 \\
I2T--vigpt2      & 0.02 & 00:00:30 \\
\midrule
I2I--sdxl        & 0.04 & 00:02:00 \\
I2I--small       & 0.01 & 00:00:30\\
\midrule
T2V--ms          & 0.005 & 03:00:00 \\
T2V--zeroscope   & 0.005 & 03:00:00 \\
\bottomrule
\end{tabular}
\end{center}
\end{table}

\begin{figure}[htbp]
    \centering
    \includegraphics[width=0.95\linewidth]{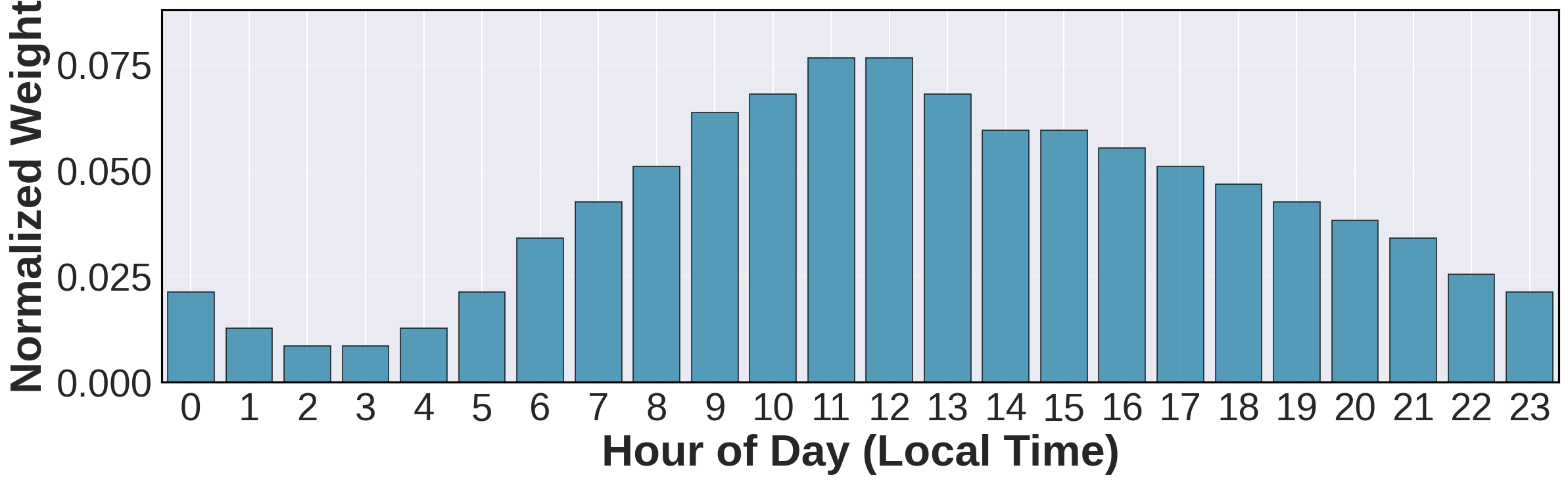}
    \caption{Diurnal arrival curve. We simulate the pattern with more tasks arriving during the day, especially around noon, and low usage after midnight.}
    \label{fig:arrival_curve}
\end{figure}

We generate task arrivals uniformly across four U.S. grid regions: CAISO North, ERCOT Austin, PJM DC, and MISO Mason City, with each region contributing one-quarter of all arrivals. Under the spatial policy, tasks may later be routed to other regions. Task arrivals within each region follow a fixed diurnal pattern in local time. Prior studies \cite{weng2022mlaas, wang2025burstgpt, xiang2025servegen} have shown that AI inference requests exhibited strong diurnal variations, with peak traffic more than twice that during off-peak hours. Our diurnal cycle, shown in Figure~\ref{fig:arrival_curve}, reflects higher traffic during daytime with a mid-day peak and lower load after midnight.

\subsection{System Configuration and Policy Settings}
We deploy the trace on a four-region GPU fleet modeled as in Section~\ref{sec:cloud_model}. Each region hosts 65 GPUs with the same heterogeneous mix: 15$\times$H100, 25$\times$A100, 10$\times$H200, and 15$\times$A6000, for a total of $G=260$ GPUs. The simulation horizon is $H=24$ hours ($86{,}400$ seconds) starting at 2024-04-13 00:00:00 UTC. Carbon-intensity signals for the four regions come from five-minute signals over that day.

We measure capacity and load using GPU-seconds. Let $G$ be the total number of GPUs in the fleet, $H$ the simulation horizon (seconds), $N$ the total number of tasks, and $\bar{r}$ denote the mix-weighted mean runtime per job, computed from the task mix and per-GPU benchmarks. For our configuration, $\bar{r}=13.70$s. The fleet capacity $C$ and trace-required load $L$ are:
\begin{equation}
    C = G \cdot H,\qquad L = N \cdot \bar{r}
\end{equation}
We define the fleet-average utilization as:
\begin{equation}
    \mu_{avg} = \frac{L}{C}
\end{equation}
To make $\bar{r}$ explicit and reusable, we first compute a per-GPU-type mix-weighted runtime:
\begin{equation}
    \bar{r}_g = \sum_{t \in \mathcal{T}} p_t \cdot r_{t,g}
\end{equation}
where $p_t$ is the task mix weight and $r_{t,g}$ is the measured mean runtime of task-model $t$ on GPU type $g$. Let $\alpha_g$ denotes the fraction of tasks that run on GPU type $g$ (in our experiment $\alpha_g$ is weighted by the GPU capacity), the fleet-level mean runtime is:
\begin{equation}
    \bar{r} = \sum_{g \in \mathcal{G}} \alpha_g \cdot \bar{r}_g
\end{equation}

The resulting fleet-average utilization is $\mu_{avg} \approx 0.37$. Because arrivals are diurnal, we also report a peak-hour utilization estimate. Using the 24-hour weight vector, the ratio of the peak hour to the average hour is $M_{peak} = 1.84$ (normalized peak hour weight over average weight), the peak utilization is $\mu_{peak} \approx M_{peak} \cdot \mu_{avg} = 0.67$.

Queuing theory has shown that system utilization must be kept well below 100\% to avoid exponentially increasing wait times, particularly in systems with high variance in request arrivals and service times~\cite{boxma1979approximations, harchol2013performance}. We therefore configure our simulation with an average utilization of $\mu_{avg} = 0.37$ and peak utilization of $\mu_{peak} = 0.67$. This conservative provisioning is necessary for the diurnal arrival pattern and the heterogeneous task mix, both of which contributing to high coefficients of variation in arrivals and service times, as observed in production LLM workloads~\cite{weng2022mlaas, xiang2025servegen}.

For both spatial and temporal shifting policies, we set the relative saving gate $\epsilon_{s,rel} = \epsilon_{t,rel} = 0.05$, which requires a $\ge 5\% $ predicted $\text{CO}_2$ reduction to shift, and the absolute saving gate $\epsilon_{s,abs} = \epsilon_{t,abs} = 0.02 (\text{gCO}_2)$. The absolute saving gate threshold is determined based on the Speed-First baseline scheduling results (avg$=0.15$g/task, p95$=0.34$g/task), calibrated to be roughly $10\%$ of the average per-task savings under the Speed-First baseline. For temporal policy guards, we set rate cap $\alpha = 0.10$, capacity ledger $\rho = 0.10$, and backlog cap $\beta=0.30$.

\subsection{Results and Analysis}
\begin{table*}[htbp]
\centering
\caption{Results across four scheduling policies}
\begin{tabular}{lrrrrrrrr}
\toprule
Policies & Total $\text{CO}_2$ & $\text{CO}_2$/task & Total Energy & Runtime/task & Queue Wait/task & SLA Violation Rate & Shifted Rate & Sch. Time \\
& (lbs) & (gram) & (MWh) & (second) & (second) & (\%) & (\%) & (ms/task) \\
\midrule
$\text{Baseline}_\text{{speed}}$  & 251.45 & 0.19 & 902.92 & 8.84 & 0.56 & - & - & 0.06 \\
$\text{Baseline}_\text{{carbon}}$ & 214.57 & 0.16 & 768.68 & 53.50 & 44.92 & - & - & 0.58 \\
Spatial  & 154.91 & 0.12 & 893.80 & 136.32 & 126.18 & - & 59.88 & 2.32 \\
Temporal & 210.83 & 0.16 & 767.94 & 66.51 & 57.92 & 3.27 & 0.4 & 0.88 \\
\bottomrule
\end{tabular}
\label{tab:result_overall}
\end{table*}

We evaluate four schedulers on the same trace at a fleet-average utilization of $\sim0.37$ with a heterogeneous GPU mix: two baselines (Speed-First and Carbon-First) and two carbon-aware policies (Spatial Shifting and Temporal Shifting).

Table~\ref{tab:result_overall} summarizes the system-level outcomes. In a heterogeneous GPU-type setting, the Carbon-First baseline lowers total $\text{CO}_2$ by 14.7\% (251.45 to 214.57 lbs) and energy by 14.9\% (902.92 to 768.68 MWh) relative to the Speed-First baseline, but at a latency cost with per-task runtime increases from 8.84 to 53.50 seconds and queue wait from 0.56 to 44.92 second. The Spatial Shifting policy achieves the lowest carbon emissions with a 38.4\% $\text{CO}_2$ reduction relative to Speed-First (241.45 to 154.91 lbs) and a 27.8\% reduction to Carbon-First (from 214.57 lbs). This comes with higher per-task latency (136.32 s runtime and 126.18 s queue wait). Total energy consumption is similar to the Speed-First baseline (-1\%, 902.92 to 893.80 MWh), but 16.3\% higher than Carbon-First, due to routing tasks to cleaner regions that may run on less-efficient GPUs. The policy shifts around 59\% of tasks across regions.

The Temporal Shifting policy delivers a 16.1\% $\text{CO}_2$ reduction compare to Speed-First (251.45 to 210.83 lbs) and is roughly on par with Carbon-First on both $\text{CO}_2$ (-1.7\%, 214.57 to 210.83 lbs) and energy consumption (-0.1\%, 768.68 to 767.94 MWh). Per-task latency is higher than under Carbon-First (+24.3\% in runtime and +28.9\% in queue wait). With the configured guards and saving gates, the SLA violation rate is 3.27\%, with a deferred fraction of 0.4\%. Scheduler overheads are negligible, for the most time-consuming policy (Spatial Shifting), with 2.32 ms/task, the overhead is less than 0.002\% of its mean runtime (136.32 seconds).

\begin{figure}[htbp]
    \centering
    \includegraphics[width=0.95\linewidth]{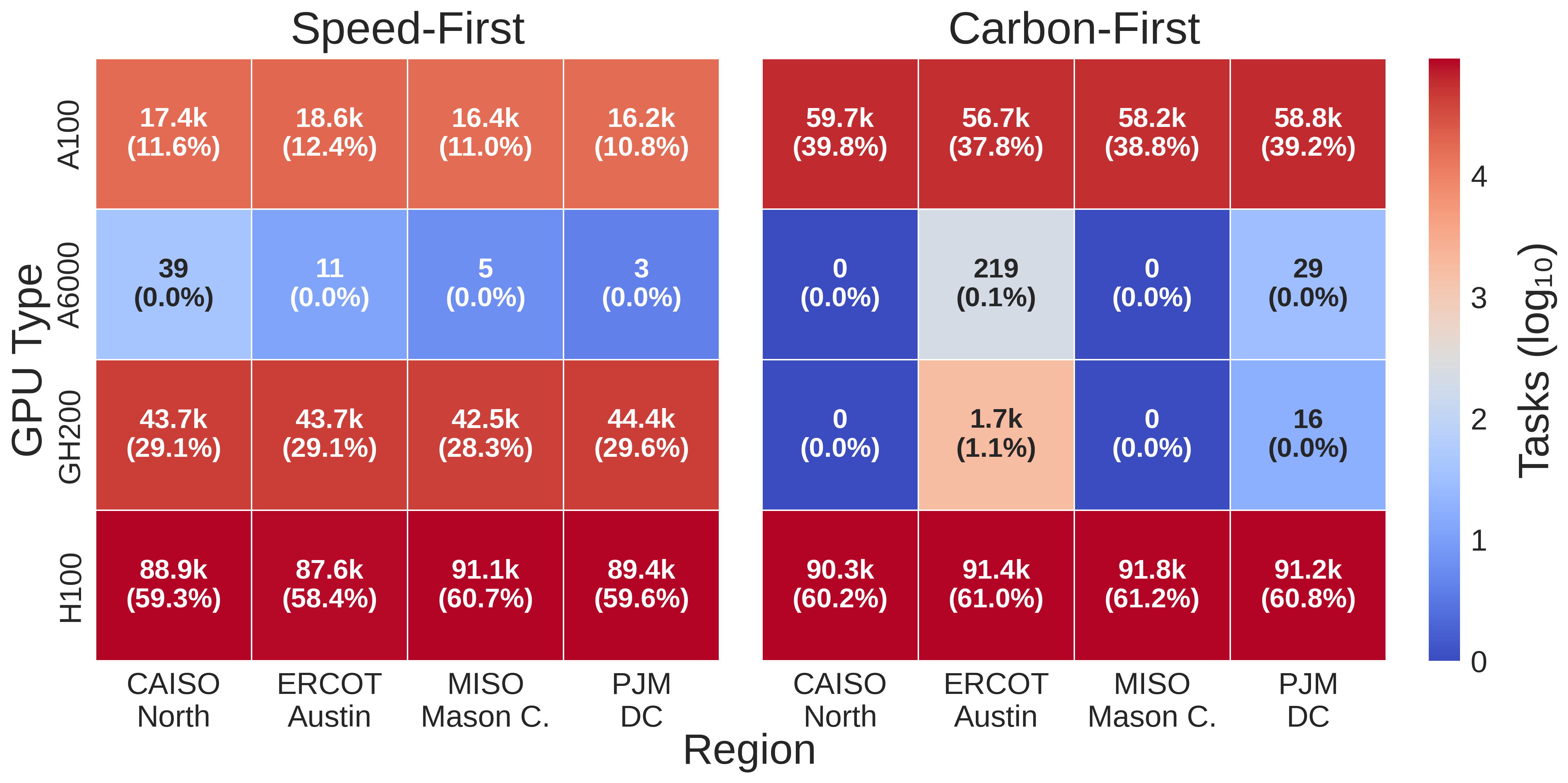}
    \caption{Task distribution across GPU type on two baseline policies. Each row represent one GPU pool and columns indicate different regions, Speed-First policy prefer faster GPUs such as H100 and GH200, while Carbon-First policy prefer energy efficient GPUs like H100 and A100.}
    \label{fig:comp_heatmap_base}
\end{figure}

When comparing the two baselines (Speed-First and Carbon-First), both execute tasks locally, the performance gap arises from GPU selection. Figure~\ref{fig:comp_heatmap_base} visualizes the task distribution across GPU types under the two baselines.
In Speed-First (left), H100s dominate (59.3\% to 60.7\% in every region), GH200s handle another large fraction (28.3\% to 29.6\%), and 10.8\% to 12.4\% of tasks run on A100s, while A6000s nearly unused.
Carbon-First redistributes toward lower-energy devices: H100 and A100 collectively handle over $99\%$ of the tasks. This device-mix shift explains the 14.9\% energy drop and 14.7\% $\text{CO}_2$ reduction in Table~\ref{tab:result_overall}.
The figure also shows that ERCOT Austin and PJM DC retain a small GH200/A6000 footprint while there is none in CAISO North and MISO Mason City. The reason is that Carbon-First policy minimizes $\text{CI}(t)\times\text{energy}$ upon task arrival. When queues on H100/A100 are long and local carbon-intensity is rising, starting immediately on a less efficient GPU can beat waiting into a dirtier interval, as ERCOT Austin and PJM DC have more volatile intraday profiles (cf. Fig.~\ref{fig:daily_moer}).
This task distribution pattern shows that on heterogeneous fleets, energy-oriented device choice alone can deliver material carbon gains, mostly by shifting tasks from GH200 to A100 while keeping H100 saturated, and that carbon-intensity volatility leads to residual use of less-efficient GPUs.

\begin{figure}[htbp]
    \centering
    \includegraphics[width=0.95\linewidth]{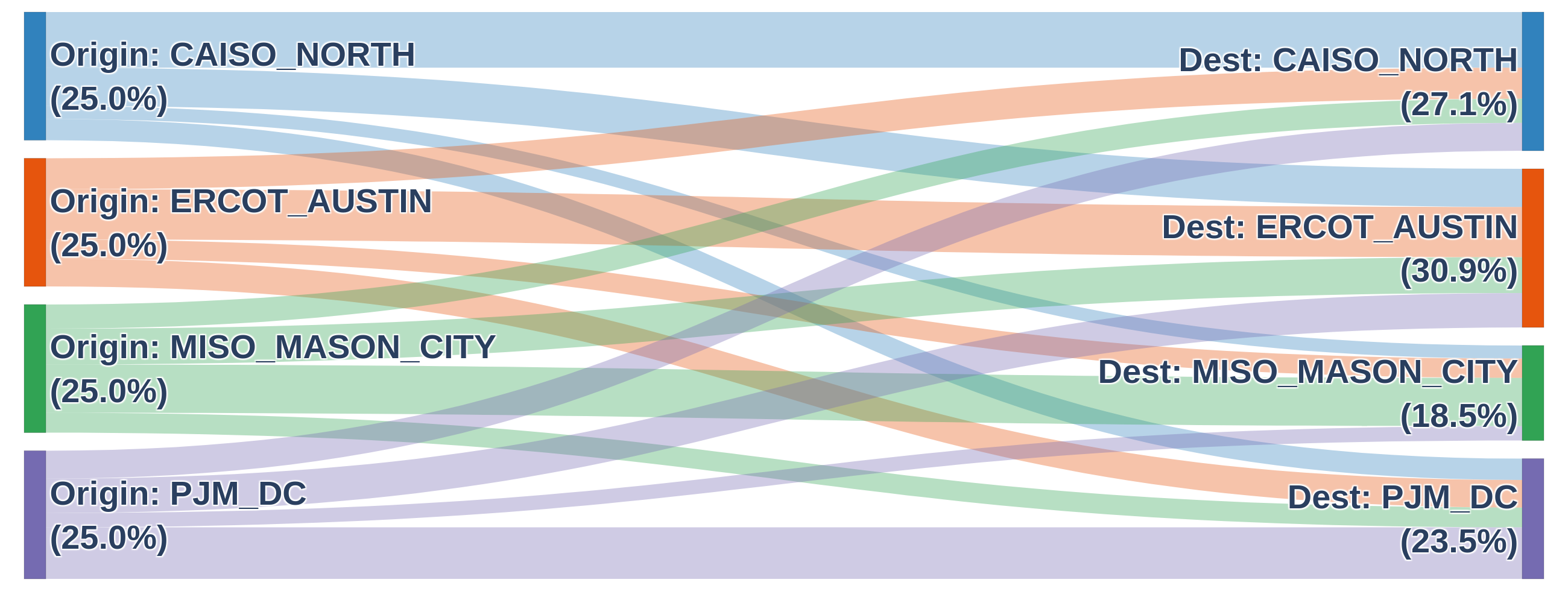}
    \caption{Spatial shift flow across regions. This Sankey diagram visualize task flow under spatial shifting policy, the left side is task original regions and right side is destination regions where task actually runs in.}
    \label{fig:spatial_shift_flow}
\end{figure}

With the Spatial Shifting policy, tasks may be executed in a region different from their origin when the destination's $\text{CI}(t) \times energy$ is lower. Figure~\ref{fig:spatial_shift_flow} visualizes the resulting flows. The trace starts with an even origin split (25\% per region). After routing, ERCOT Austin and CAISO North become net importers, executing 30.9\% and 27.1\% of all tasks, respectively, while PJM DC and MISO Mason City are net exporters, ending at 23.5\% and 18.5\%. This rebalancing is consistent with their carbon-intensity profiles on the study day, where ERCOT Austin and CAISO North have, on average, lower carbon-intensity than the other regions and ERCOT Austin shows several intraday troughs, so directing arrivals there yields the 38.4\% total $\text{CO}_2$ reduction reported earlier.
Notably the flow pattern is many-to-many: each origin sends work to all four destinations and each destination receives work from all four origins, this breadth comes from the interleaving carbon-intensity over time.

\begin{figure}[htbp]
    \centering
    \includegraphics[width=1.0\linewidth]{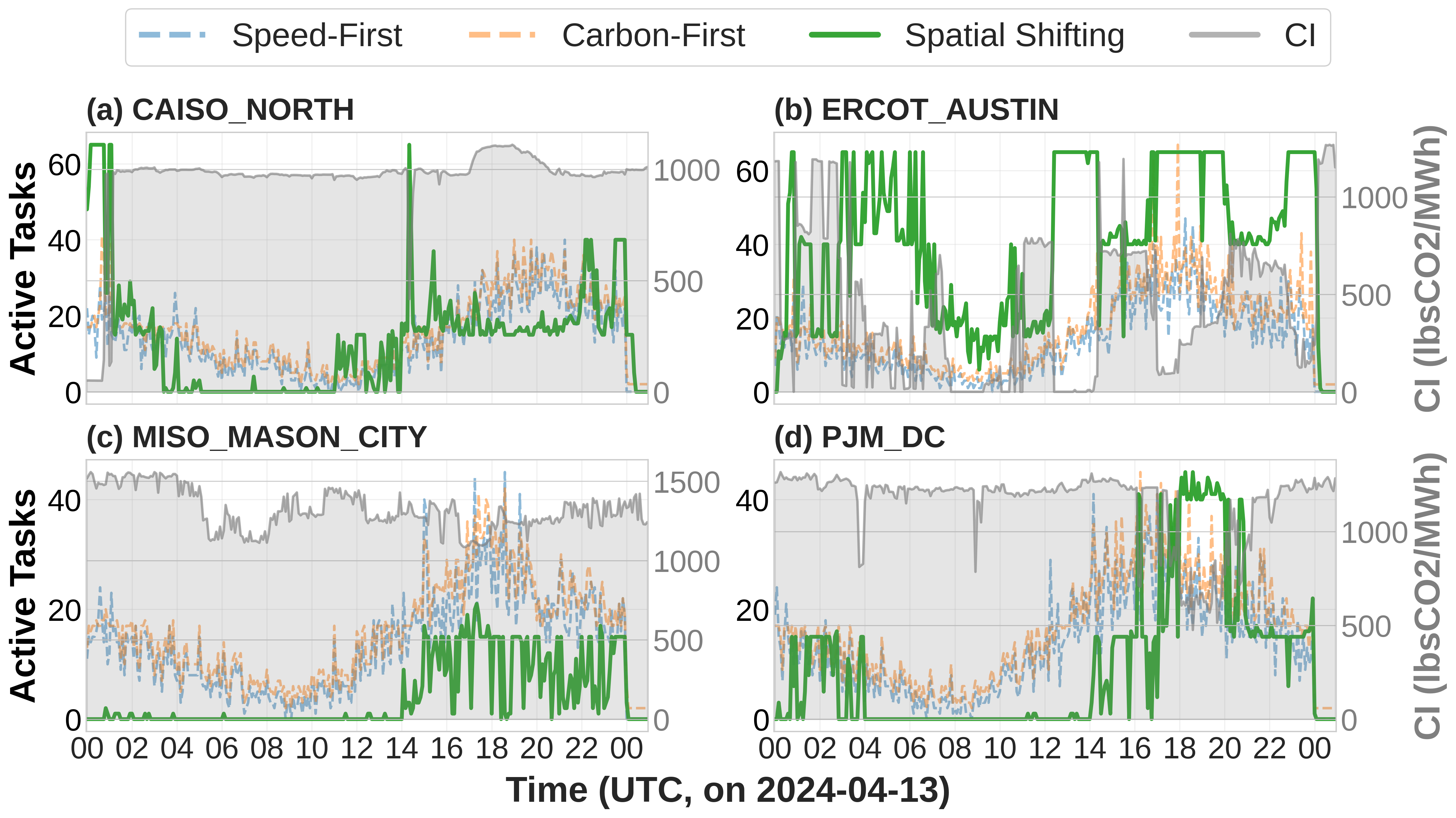}
    \caption{Spatial shift scheduling in a 24 hour window across four regions. Each plot shows the number of active tasks under three shifting policies (blue dashed: Speed-First, orange dashed: Carbon-First, green solid: Spatial Shifting) and the carbon-intensity (gray line with filled color). Two baseline policies follow task arrival pattern while spatial shifting policy schedule more tasks at low-CI window, capped by GPU capacity (65 GPUs per region).}
    \label{fig:spatial_shift_active_jobs}
\end{figure}

Figure~\ref{fig:spatial_shift_active_jobs} plots the number of active tasks (left axis) and each region's carbon-intensity (right axis). The dashed baselines (Speed-First, Carbon-First) largely follow the diurnal arrival pattern. In contrast, the Spatial policy (solid green) tends to anti-correlates with carbon-intensity: it ramps up when the local grid is cleaner and backs off when it is dirtier.
In CAISO North (panel a), carbon-intensity is lowest near 00-01 UTC; the Spatial policy immediately drives the region to its capacity limit (65 concurrent tasks). In ERCOT Austin (panel b), extended low-CI intervals appear around 04-06, 12-14, 17-20, and 22-23 UTC; during these windows, the Spatial curve reach close to the capacity limit, indicating continuous full utilization when ERCOT is the cleanest option. This behavior is consistent with the policy objective.
From 01-04 UTC in PJM DC (panel d) and from 16-23 UTC for both CAISO North (panel a) and MISO Mason City (panel c), we observe non-zero Spatial activity even when their CI is relatively high. This reflects regional capacity limits: when the cleaner regions (e.g. ERCOT, sometimes CAISO) are saturated yet demand remains high, the scheduler must place some tasks in the next-beset regions to meet SLA constraints.

Under temporal shifting, tasks may be deferred to a later start time within the same region when doing so lowers carbon emission while respecting the task's delay limit. In our configuration, this yields an additional 1.7\% $\text{CO}_2$ reduction over the Carbon-First baseline, by deferring 0.4\% of tasks (2,427 of 600,000). Unlike the other policies, temporal scheduling introduces SLA breaches. These SLA violations happen because deferred tasks in the priority queue and non-deferred tasks in the normal queue contend for the GPU pool.
In our run, the SLA-violation rate is 3.27\% (19,602 tasks): 97 violations are deferred tasks that slipped behind normal tasks, and 19,505 are non-deferred tasks that were further delayed when priority tasks overtook their place.

\begin{figure}[htbp]
    \centering
    \includegraphics[width=1.0\linewidth]{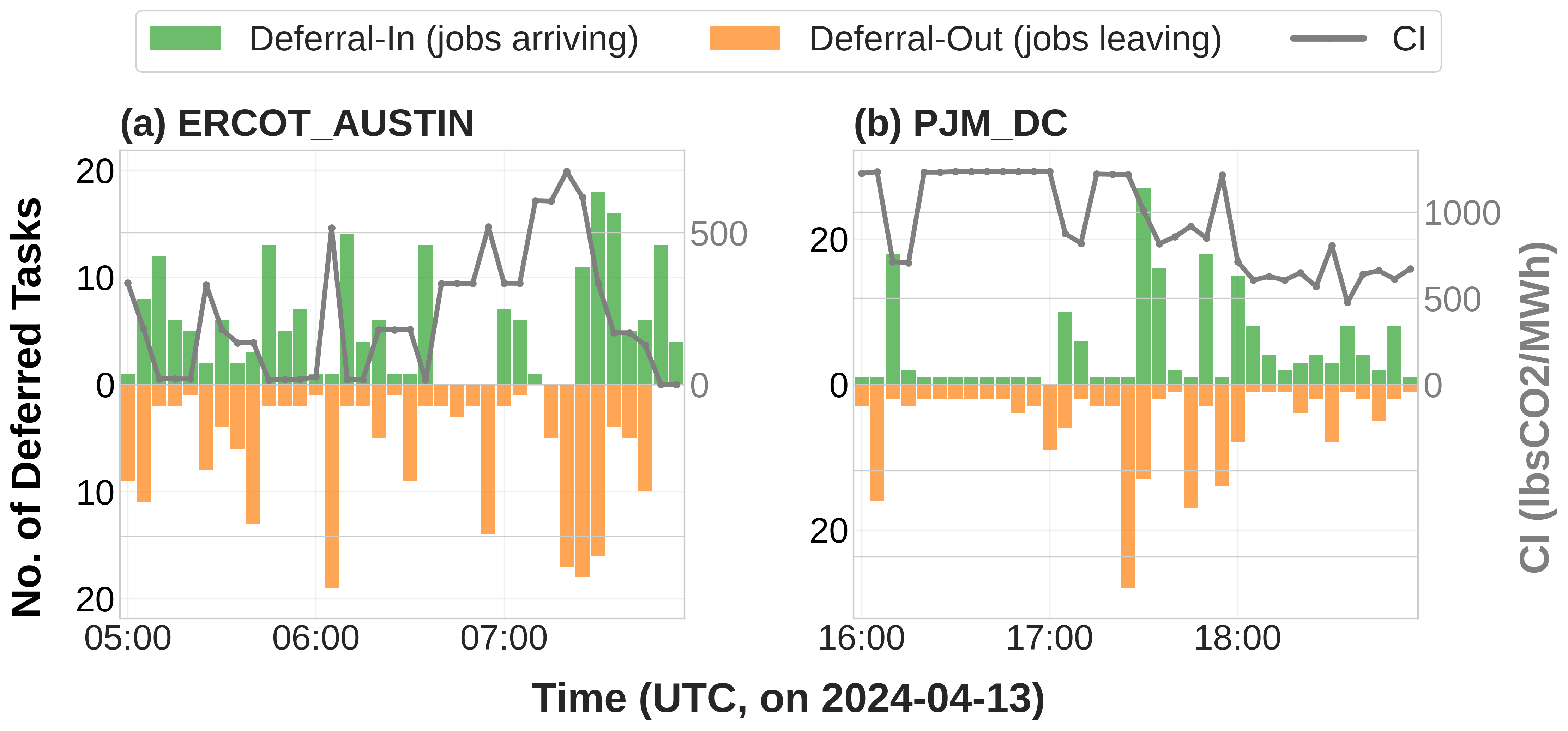}
    \caption{Temporal shifting detailed on two regions. Each plot shows the number of deferred tasks (green bars: in, orange bars: out) and the carbon-intensity (gray line) in each 5-minute window.}
    \label{fig:temporal_shift_defer_jobs}
\end{figure}
Figure~\ref{fig:temporal_shift_defer_jobs} visualizes five-minute-bucket deferral activities for two regions with significant carbon-intensity variability. Bars above zero (Deferral-In) count tasks that begin in the bucket due to earlier deferrals; bars below zero (Deferral-Out) count tasks deferred out of the bucket at arrival. The blue line shows the region's carbon-intensity. When carbon-intensity drops between consecutive five-minute intervals, we consistently see a surge of Deferral-Out in the higher-CI bucket followed by a surge of Deferral-In in the next lower-CI bucket. This pattern is visible in both ERCOT Austin and PJM DC and is precisely the behavior the temporal policy is designed to induce: shift a portion of arriving tasks across near-term buckets to catch cleaner windows without overwhelming future capacity.
Temporal shifting is effective only under strict conditions. First, it needs frequent, sizable CI swings. When the grid is flat, deferring adds queuing risk with trivial carbon benefit. Second, it requires sufficient slack, short, interactive tasks with tight delay limits rarely benefit. Third, it is operationally complex because it depends on carefully tuned guard thresholds, and contention between the priority and normal queues can cause violations.

Decision cost scales with search scope and carbon accounting, but remains negligible. As shown in Table VI, Speed-First inspects only the origin's GPU pools without carbon accounting (0.06 ms/task; 34.9 s total for 600k tasks). Carbon-First stays local and evaluates predicted $\text{CO}_2$ for each GPU type (0.58 ms/task; 346 s). Spatial expands the candidate set to all regions, computing $\text{CO}_2$ per region-GPU pair (2.32 ms/task; 1,394 s), which corresponds roughly to a 4$\times$ increase over the local-only policies in our case. Temporal remains local but scans near-term windows up to each task's delay limit (0.88 ms/task; 531 s). Even the slowest policy accounts for less than 0.002\% of mean service time. We therefore treat scheduling time as trivial and exclude it from latency/energy/$\text{CO}_2$ comparisons.

To isolate routing effects from device heterogeneity, we repeat the study with four regions each hosting an identical pool of 65$\times$A100 GPUs, keeping the workload, diurnal curve, delay limits, and scheduling thresholds unchanged. The results are listed in Table~\ref{tab:single_gpu_result}.
\begin{table}[htbp]
\caption{Heterogeneous vs Single GPU test results}
\centering
\begin{tabular}{lrrr}
\toprule
GPU pool/region & Policy & Carbon (lbs) & Energy (MWh) \\
\midrule
\multirow{4}{*}{\begin{tabular}[c]{@{}l@{}}25xA100, 15xA6000\\ 15xH100, 10xGH200\end{tabular}}
 & Speed-First & 251.45 & 902.92 \\
 & Carbon-First & 214.57 & 768.68 \\
 & Spatial  & 154.91    & 893.80    \\
 & Temporal & 210.83    & 767.94    \\
\midrule
\multirow{4}{*}{65xA100}
 & Speed-First & 283.38 & \multirow{4}{*}{1,007.83} \\
 & Carbon-First & 283.38 \\
 & Spatial  & 205.35 & \\
 & Temporal & 277.35 & \\
\bottomrule
\end{tabular}
\label{tab:single_gpu_result}
\end{table}
Under this single-GPU configuration, every task runs on the same device, total energy is identical across policies, and the two baseline policies become the same. The energy increase observed in heterogeneous fleet stems from GPU mix: when the scheduler minimizes $\text{CO}_2$, it sometimes choose cleaner grids on less energy-efficient GPUs. The carbon benefits persist: the Spatial policy lowers $\text{CO}_2$ by 27.5\% relative to the baseline, while the Temporal policy yields a $\sim2\%$ reduction. These results suggest that energy increases appear when device efficiency varies and can be constrained by an energy-aware strategy if desired.

\section{Conclusions and Future Work}
In this paper, we propose to reduce carbon emissions of AI data centers via carbon-aware siting and scheduling. Using TeleGeography locations aligned with high-resolution marginal carbon-intensity data, we visualize and quantify siting-grid misalignment across the U.S., EU (incl. UK), and Australia and observed that existing and recent built data centers are not preferentially placed in low-carbon-intensity regions. We then design the Carbon-Aware Task Simulator (CATS) and use it to quantitatively evaluate the impact of spatial and temporal shifting on carbon emission reduction.

Our results show that spatial shifting yields significant carbon savings: with around 60\% tasks shifted, it reduces $\text{CO}_2$ by 28\% relative to carbon-first and 38\% over speed-first. Temporal shifting reduces 1.7\% versus carbon-first and 16.1\% against speed-first, deferring around 0.4\% of total tasks, but introduced 3.3\% SLA violations due to task contentions. Both policies have minimal scheduling overhead (a few milliseconds per task).

This work has several limitations that suggest clear directions for future research. First, it relies on historical carbon-intensity data; we plan to incorporate short-term forecasts and evaluate robustness to forecast errors. Second, we omit financial costs; a multi-objective formulation that jointly optimize carbon, energy, latency, and financial cost is a natural extension. Third, we assume fixed GPU capacity; enabling elastic capacity in response to arrival patterns is another important next step.

\bibliography{reference}
\bibliographystyle{IEEEtran}

\begin{IEEEbiography}[{\includegraphics[width=1in,height=1.25in,clip,keepaspectratio]{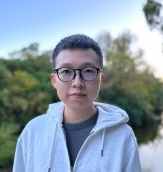}}]{Dayuan Chen} received his M.S. in Computer Science from University of Texas at Dallas in 2021. He is now a Ph.D. student at Texas State University. His research focuses on sustainable AI development, including efficient and carbon-aware Large Models fine-tuning and inference scheduling.
\end{IEEEbiography}
\begin{IEEEbiography}[{\includegraphics[width=1in,height=1.25in,clip,keepaspectratio]{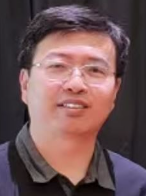}}]{Ziliang Zong} received his Ph.D. degree from Auburn University and is a Professor of the Computer Science Department at Texas State University. His research focuses on Energy-Efficient Computing and Systems, including Green Software Design, Green AI, Green Cloud, and Green Data Center. He served as a member of the Green Software Foundation, Associate Editor of the Sustainable Computing Journal, co-chairs and committee members of numerous conferences and workshops in high performance computing, green computing, cloud computing, and edge computing.
\end{IEEEbiography}

\end{document}